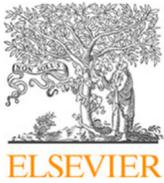
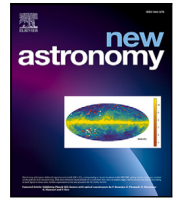

# Investigating star formation scenarios in the Milky Way using YSO distributions—A pilot study

Annarien G. Headley [a],*, James O. Chibueze [a,b]

[a] *UNISA Centre for Astrophysics and Space Sciences (UCASS), College of Science, Engineering and Technology, University of South Africa, Cnr Christian de Wet Rd and Pioneer Avenue, Florida 1709, P.O. Box 392, 0003 UNISA, South Africa*
[b] *Department of Physics and Astronomy, Faculty of Physical Sciences, University of Nigeria, Carver Building, 1 University Road, Nsukka 410001, Nigeria*

ABSTRACT

We investigated the distributions of classified young stellar object (YSO) in nine star-forming regions associated with H$_{II}$ regions, namely Sh2-22, Sh2-19, Sh2-17, M16, M8, IC5070, Sh2-252, NGC2467 and M42, as a means of exploring the star formation scenarios (triggered or spontaneous) in the various regions. The YSO distributions of nine regions along the galactic longitudes and across different spiral arms were explored. It is noted that Sh2-22 and Sh2-19 may have evidence of localized triggered star formation, whereas Sh2-17 may be a spontaneous star forming region. The results showed that the location within the Milky Way may influence localized triggered star formation within individual regions. There is strong evidence that there are fewer older stars (Transition Disks) within the Outer regions of the Galaxy ($2.91 \pm 1.74\%$), accompanied by a higher number of Class I ($32.49 \pm 10.77\%$). This is different from the Inner Galactic regions where the Transition Disks are higher ($32.93 \pm 13.78\%$) as compared to low Class I sources ($11.01 \pm 4.81\%$). The number of Class II stars is relatively high and increases from the Inner regions ($52.64 \pm 10.07\%$) to Outer regions ($64.54 \pm 11.09\%$). These nine star-forming regions are not an accurate representation of locations throughout the Galaxy and the results may be biased.

## 1. Introduction

Stars in the Milky Way are formed in Giant Molecular Clouds (GMCs). GMCs containing high-mass OB stars are associated with ionized hydrogen H$_{II}$ regions because of energetic ultraviolet (UV) photons released by OB stars. Stars in these GMCs are formed mainly through two processes, spontaneous star formation and triggered star formation (sometimes referred to as sequential star formation). Spontaneous star formation occurs when a dense core inside a dense molecular cloud collapses under its own gravity without external factors (Kerton et al., 2008; Kinoshita et al., 2021). Triggered (or sequential) star formation is caused when ionization from OB stars compresses dense cloud cores, inducing star formation. OB stars have profound impacts on the surrounding molecular clouds in which they are formed. These OB stars cause ionization and expansion of ionized hydrogen in the form of H$_{II}$ regions. The expanding H$_{II}$ regions along with the kinetic energy from strong stellar winds and shocks from supernovae compress or disperse nearby clouds. As this compression front sweeps over the molecular cloud, dense cores collapse, triggering the formation of new stars (Deharveng et al., 2005; Bieging et al., 2009; Elmegreen, 2011; Li et al., 2014; Kinoshita et al., 2021).

High mass stars such as OB (M $\gtrsim 8 - 12 M_\odot$) stars are often used when studying star formation (Sabatini et al., 2021), these stars have relatively short lifetimes and often end in violent supernova explosions producing new heavier elements into their surroundings. The typical time scales for these high mass OB type stars is around 1-20 Myrs (Blaauw, 1964; Wright, 2020; Ge et al., 2024). In OB associations, the intense ionization front destroys the immediate surrounding molecular cloud. As the distance from the OB stars increases and the speed of the ionization front decreases, the ionization front sweeps over the neighboring molecular cloud cores triggering the formation of new stars (Preibisch and Zinnecker, 2007). The common evidence of triggered star formation is: (1) there is an OB association source responsible for the expanding ionization front of the H$_{II}$ region, (2) OB stars are often older than the stars that were formed along the ionization front, (3) an age gradient of young stellar objects (YSOs) is seen. Here older stars (Class II) are formed along the ionization front and are the closest to the OB stars followed by the younger stars (Class 0/I) in the densest regions of the cloud (Preibisch and Zinnecker, 2007). As the H$_{II}$ region of a high-mass star expands into the surrounding interstellar medium, gas is accumulated and compressed as an ionization front sweeps over the surrounding molecular cloud. The dense cores of gas eventually gravitationally collapses, forming new stars (Deharveng et al., 2005; Elmegreen, 2011; Li et al., 2014). There is usually stability in the molecular cloud before an ionization front arises and instability after the ionization front has passed over the cloud; this phenomenon is often associated with triggering (Elmegreen, 2011). As the H<sub>II</sub> regions expand into nearby molecular clouds, the low-density material dissipates faster than denser cores. This reveals structures such as bright rims and pillars within the cloud (Elmegreen, 2011).

In a triggered region, it is common to see "onion-like" layering: The oldest Transition Disks are located throughout the cloud. The old Class II stars are seen along the ionization front where the interaction between the expanding H<sub>II</sub> region and the molecular cloud first took place. The Class I stars are located deeper within the molecular cloud, followed by the youngest stars within the densest cores embedded in the cloud. There are not many young stars located deep inside the cloud, rather most are located in dense cloud cores along the ionization front (Chen et al., 2007; Choudhury et al., 2010; Elmegreen, 2011; Chibueze et al., 2012).

Star formation within galaxies is governed by the local star formation within molecular clouds. The rate at which the molecular cloud gas is transformed into a new protostar is dependent on the mass of the clouds dense gas (Lada et al., 2012). The rate of star formation within a galaxy is determined by local processes within the clouds rather than larger galactic processes. However, galactic processes may contribute to the amount of dense gas within the local molecular cloud environment (Lada et al., 2012). The longitude-velocity plots of Urquhart et al. (2014) in Fig. 1 shows the structure of the Milky Way Galaxy along the Galactic Plane. Within the inner Galaxy at longitudes $|l| < 60°$, a strong relationship between the molecular gas and the high mass stars from the Red MSX Source (RMS) survey is seen. However, at $|l| > 60°$ this relationship between molecular gas and the number of star-forming regions declines substantially. Within this image the estimated positions of the galactic spiral arms are also revealed. There is a strong relationship between the number of H<sub>II</sub> regions and subsequent high mass young stars with the locations of the spiral arms. This is especially true in the outskirts of the Galactic Plane, where the molecular gas is less prominent (Urquhart et al., 2014).

Several studies have been performed to determine star formation in the H<sub>II</sub> regions, but these have resulted in conflicting results as the triggering of star formation in these areas is not fully understood. An example of this is in M16, where the distribution of stellar objects shows that Type I objects are present in the pillar heads of the molecular clouds and that older objects are randomly located in the region (Sugitani et al., 2002). However, in a later study the young protostars were observed to be randomly located in the region with no clear age pattern (Indebetouw et al., 2007). Previous studies of M16 have shown that it is difficult to determine whether star formation in dense molecular clouds was triggered by interactions with the H<sub>II</sub> regions or if stars were formed in the pillar heads before the pressure of ionized hydrogen from the H<sub>II</sub> regions (Elmegreen, 2011).

Another example is IC1396, where several Class I stars have been observed in the pillars, and Class II stars are scattered over the region. The Class I stars may have been triggered as they are younger than the H<sub>II</sub> region (Elmegreen, 2011). Although in a study conducted by Beltrán et al. (2009) of the same area (IC1396N), no color gradients, age gradients, or patterns indicating triggered star formation were observed. However, in a later study, when the same area was observed using the Spitzer Infrared Array Camera (IRAC) and the Multi-band Imaging Photometer (MIP), an age gradient was observed (Choudhury et al., 2010). This study showed that both an O6.5 star (HD20267) and a B0V star (H206773) are responsible for the ionization of IC1396N. Here an age gradient from South to North was observed, with Class II stars closest to the ionization front, followed by Class I and 0 stars deeply embedded within the dense cores of the cloud (Choudhury et al., 2010). This follows the onion-like layering expected for triggered regions where the oldest stars (in this case Class II stars) are formed first as the compression front from the ionization interacts with the molecular cloud.

In these previous studies, there was no clear consistency in the correlation between when and where triggered star formation occurs and the circumstances that governs the formation of new stars. Each study uses a different classification scheme and data, and looks at different molecular clouds. These differences may affect the resulting variations in the results.

The aim of this pilot study is to determine if the distribution of local triggered star formation within a star-forming region associated with an H<sub>II</sub> region changes as the location within the Milky Way Galaxy changes. To overcome the inconsistencies seen in previous studies between molecular clouds, the same survey data is used (2012 data release of Wide-field Infrared Survey Explorer Cutri et al., 2012), as well as the same classification scheme. This study might also give insights into how the location of a molecular cloud within the Galaxy influences the local star formation in the cloud. The investigations of this pilot study are completed by studying the variations of YSO distributions of Class I, Class II and Transition Disks within nine star-forming regions. This will be analyzed in two different ways: the first is to analyze the YSO distributions according to galactic longitudes. The second is to analyze the YSO distributions according to location in the spiral arms. This pilot study will provide the foundation for future investigations to explore triggering in localized environments throughout the Milky Way Galaxy in a similar manner.

In Section 2, the methodology for gaining the YSO distributions is discussed. In Section 3, the YSO distributions are presented for each of the nine star-forming regions. In Section 4, the results are analyzed in terms of the galactic longitudes and the spiral arms. In Section 5 is the summary and conclusion of the key findings made in this paper.

## 2. Method

This pilot study tries to determine how triggered star formation changes throughout the Milky Way Galaxy. We have divided the Galactic Plane into three separate groups: the Inner, Middle and Outer Groups as seen in Fig. 1. This will allow us to group together star-forming regions that may have similar velocities along the galactic plane (Urquhart et al., 2014). The locations of the H<sub>II</sub> regions are presented in the longitude-velocity graphs of Urquhart et al. (2014) and in Fig. 1.

### 2.1. Selection criteria of star-forming regions

The selection criteria for the nine regions in this pilot study is that there is known star formation within the molecular clouds and that there is evidence of triggered star formation. The galactic longitudes have been divided into three separate groups: the Inner, Middle, and Outer Groups as seen in Fig. 1. The Inner Group has a galactic longitude range of $355° - 5°$. The Middle Group has galactic longitude ranges of $5° - 90°$ and $270° - 355°$. The Outer Group has a galactic longitude range of $90° - 270°$. Within each of these groups, three different regions with molecular cloud structures and active star formation were chosen to represent the groups.

These regions in each group were selected as there is visible signs of star formation and has an O/B type stars nearby. These massive stars drives the expansion of ionized hydrogen within the gas cloud, triggering of the formation of new stars within the cloud. The regions Sh2-22, Sh2-17 and Sh-19 are of particular interest as very little previous studies were done on these star-forming regions. The regions chosen for the Inner Group is Sh-22, Sh-19, and Sh-17. The Middle Group includes the Lagoon Nebula (M8), the Pelican Nebula (IC5070) and the Eagle Nebula (M16). The Outer Group includes the Monkey Head Nebula (Sh-252), the Skull and Crossbones Nebula (NGC2467), and the Orion Nebula (M42). The literature insights for each of these region are presented in the following Sections 2.1.1–2.1.9. In Fig. 1, the Inner and Middle Group do not have selected regions in the Southern Plane, as this was selected in the Northern Plane. Similarly the Outer Group does not have selected regions in the Northern Plane, as these were selected in the Southern Plane.

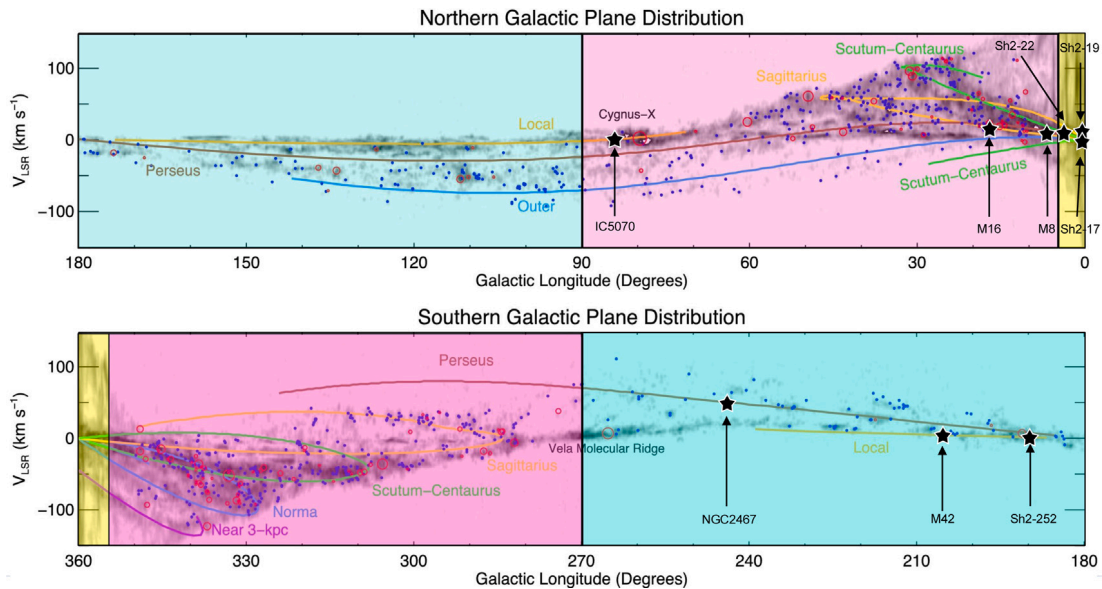

**Fig. 1.** The velocity distribution of high mass stars from the RMS against the longitudes of the Northern (top panel) and Southern (bottom panel) Galactic Planes. The red circles indicate star forming regions and their size indicates the density. The blue circles show the rest of the RMS sample data. The molecular gas of the Galaxy is detected through integrated $^{12}$CO emission. The locations of the different galactic spiral arms is indicated with the solid lines. This image was taken from Urquhart et al. (2014). For the purposes of this study, the galactic longitudes is divided into three segments indicating the locations of the Inner (yellow), Middle (yellow) and Outer (blue) Groups. The locations of the nine H II regions that is studied in this paper is shown by the black stars with white borders.

### 2.1.1. Sh2-22

Sh2-22 (RCW144, Gum71 and LBN14) is located at $(l, b) = (4.745°, -0.500°)$ (Lozinskaya et al., 1983; Megier et al., 2009). This region has not undergone extensive studies previously; only one previous study by Lozinskaya et al. (1983) could be found. In Lozinskaya et al. (1983), observations were completed using the Palomar Observatory Sky Survey taken with 48-inch Oschin Schmidt Telescope. These observations were focused on Sh2-22 and on the ionizing Of-type star HD162978 located at $(\alpha, \delta)_{J2000} = (268.731°, -24.887°)$ (Lozinskaya et al., 1983). The photometric distance to H162978 is 1.4 kpc, and is situated at 7 pc above the Galactic Plane. The mean radial velocity is $+7.5$ km s$^{-1}$ with a kinematic distance of $2.7 \pm 1.6$ kpc (Lozinskaya et al., 1983). According to Megier et al. (2009), the distance taken using Ca II parallax measurements is $799\pm107$ pc. In 1983 it was suggested that the structure of the nebula is formed through strong radiation, stellar winds and possible supernova explosions as a result of stars with OB association (Lozinskaya et al., 1983). HD162978 has evidence of weak periodic changes. There is strong evidence that there is an envelope surrounding HD162978 expanding at a rate of $1-5\times10^5$ per year (Lozinskaya et al., 1983). This may lead to possible triggered star formation within this region as a result of the Of-type star HD162978. The star produces solar winds strong enough to compress nearby molecular cloud cores and subsequently form new stars. According to the Sharpless Catalog, Sh2-22 is an H II region of moderate surface brightness and is an irregular shape with a clumpy structure in the molecular gas (Sharpless, 1959). The Ca II column density for HD162978 is $3.15\pm0.43\times10^{12}$ cm$^{-2}$ (Megier et al., 2009).

### 2.1.2. Sh2-19

Sh2-19 (RCW140, LBN2) is located close to the Galactic Center in terms of galactic longitudes. The known galactic coordinates of Sh2-19 is $(l, b) = (0.11°, -0.56°)$ from the 1989 study of Avedisova and Palous (1989). This also recorded a distance of $1.7 \pm 0.4$ kpc with a radial velocity of $+18.1$ km s$^{-1}$ (Avedisova and Palous, 1989). According to the Sharpless Catalog, Sh2-19 has a circular shape, an amorphous structure, and moderate surface brightness (Sharpless, 1959). This H II region has not been well studied, and the majority of the information found is part of previous catalogs. This H II region was selected to be part of this study after completing the YSO classification in Section 2.2 and finding that there may be evidence of triggered star formation.

### 2.1.3. Sh2-17

The H II region of Sh2-17 (RCW138) is identified in the Sharpless catalog as an irregular nebula with clumpy structure and moderate surface brightness (Sharpless, 1959). This H II region is close to the Galactic Center located at $(l, b) = (0.489°, -0.669°)$ and has an estimated diameter of $25'$ (Sharpless, 1959). This region H II has not been well studied, and there is little literature on this source.

### 2.1.4. Lagoon Nebula (M8)

The Lagoon Nebula (M8, Sh2-25, RCW146, LBN25) is a well studied star-forming region in the Sagittarius arm of the Milky Way Galaxy at a distance of 1.35 kpc $\pm9\%$ (Damiani et al., 2019). M8 is a triggered high-mass star-forming region that contains several O-type exciting stars. Within the nebula there are two sites that have strong evidence for active star formation; this includes M8-East and M8-Main as mentioned in Tiwari et al. (2020). M8-East has a higher abundance of chemical traits than those observed in M8-Main, and the star formation is deep within the dense cores of the cloud. M8-Main has a warmer and dispersed photodissociation region. The M8-East region is younger than M8-Main and has evidence of younger star formation driven by a very bright infrared source called M8E-IR located at $(\alpha, \delta)_{J2000} = (271.222°, -24.445°)$ (Tothill et al., 2008; Tiwari et al., 2020). M8-East was formed as a result of an ionization front pushing material from the central region of the nebula outwards. As this ionization front expands, compression occurs and subsequently the closest regions of the cloud is heated resulting in a bright rimmed structure of the cloud. In Tiwari et al. (2020), the YSO distribution was examined and strong evidence of triggered star formation was seen near M8-East, where the Class 0/I stars are located among the densest gas and the Class

II stars are located along the ionization front (Tiwari et al., 2020). There are a few other high mass stars aiding in the ionization of this molecular cloud. This includes HD165052 (spectroscopic binary O6.5 V star) located at $(\alpha, \delta)_{J2000} = (271.294°, -24.399°)$, 9 Sagittarii (O4 V star) at $(\alpha, \delta)_{J2000} = (270.969°, -24.361°)$, and Herschel 36 (O8V+ star) at $(\alpha, \delta)_{J2000} = (270.918°, -24.379°)$ (Tothill et al., 2008). In 2024, Kahle et al. (2024), showed that the Lagoon Nebula is a chemically complex cloud with 70 molecular traits observed. In the nebula, 38% of molecular clumps have a strong possibility of containing a protostellar object where both high and low mass star formation may occur (Kahle et al., 2024).

### 2.1.5. Pelican Nebula (IC5070)

The Pelican Nebula (IC5070, LBN350) is a well-studied star forming region. The Pelican Nebula along with the North American Nebula (NAN) is part of a larger molecular cloud complex called the NAN complex (Das et al., 2021). Within IC5070 there are traits of variability of new YSOs in terms of brightness, periodicity and light curves (Hillenbrand et al., 2022), this allows for the further study of the evolution of individual stars within the nebula (Froebrich et al., 2018; Sicilia-Aguilar et al., 2024; Herbert et al., 2024). Evidence for triggered star formation in IC5070 was found when measuring the kinematics of many young stars in this region (Fang et al., 2020) The YSO distribution showed that older stars (1–10 Myrs) are evenly distributed, however, the younger stars (<0.5 Myrs) are located within the dense molecular cores along the boundary of a H II bubble (Fang et al., 2020). New young stars are visible along the ridge due to strong radiation winds and expanding H II regions causing the dissipation of the molecular cloud surrounding these young protostars (Panwar et al., 2023). There are two well-known O-type stars that are linked to the NAN complex, these are 2MASSJ20555125+4352246 (a O3.5IIIf+O8 star at $(\alpha, \delta)_{J2000} = (313.964°, +43.873°)$) and HD199579 (a O6.5 V star at $(\alpha, \delta)_{J2000} = (314.145°, +44.925°)$) (Maíz Apellániz et al., 2016; Kuhn et al., 2020; Panwar et al., 2023).

### 2.1.6. Eagle Nebula (M16)

The Eagle Nebula (M16, NGC6611, IC4703, RCW165, LBN67) is a well-known active star forming region located in the Sagittarius arm at a distance of ∼1750 pc. M16 is a source that has been subjected to scrutiny on whether triggered star formation is present, as seen in the review study done by Elmegreen (2011). In particular the point of contention is regarding whether the stars within the pillars are made by triggering or if they were made within the dense pillar regions before being revealed as the gas dissipates revealing dense structures such as the pillars (Elmegreen, 2011). In 2002, an age gradient was detected as Type I stars were detected close to the pillar heads surrounded by older stars suggesting triggered star formation (Sugitani et al., 2002). However, in 2007, the YSOs were observed to be evenly distributed and no age gradient was detected, thus triggering was not confirmed. However, Guarcello et al. (2010) showed that triggered star formation was not present on a large-scale in the nebula but rather in a south-east to north-west direction caused by a collision with a GMC around 3 Myrs ago. This GMC was created ∼6 Myrs ago through supernovae explosions. There is little evidence to confirm that OB stars are responsible for large scale ionization and triggering (Guarcello et al., 2010). This result was supported in 2017, when CO observations were completed of the extended M16 GMC which includes a nearby bubble called N19 and NGC6611 (Nishimura et al., 2017). The M16 GMC is ∼$1.3 \times 10^5$ M$_\odot$ with a size of 20 × 30pc. This GMC has three velocity components - blue-shifted, main and red-shifted. The GMC is contains 52 OB stars, and a decrease in molecular gas is seen towards the O-type stars as a result of photo-ionization. The collision between the blue-shifted and red-shifted velocity components of the GMC produces the formation of triggered O-type stars (Nishimura et al., 2017). These findings are supported by Xu et al. (2019), who also claim that triggered star formation took place within M16 due to an expanding H II region. There are three O-type stars that may be responsible for the ionizing in this nebula. This is HD168075 (a O6.5V(f) star at $(\alpha, \delta)_{J2000} = (274.650°, -13.793°)$), HD168137 (a O8Vz star at $(\alpha, \delta)_{J2000} = (274.734°, -13.809°)$) and NGC6611-222 (O7V(f)z star at $(\alpha, \delta)_{J2000} = (274.656°, -13.728°)$)

### 2.1.7. Monkey Head Nebula (Sh2-252)

The Monkey Head Nebula (Sh2-252, NGC2174) is an optically bright H II region part of the Gemini OB1 association and is ionized by a central O6.V star. In 1995, it was suggested that the expanding H II region is responsible for the formation of embedded stars (Carpenter et al., 1995). The majority of young stars are located near compact H II regions called Sh2-252 A, C and E, indicating that there may be local triggered star formation (Jose et al., 2012). Chibueze et al. (2012) investigated the NH$_3$ emission lines measuring the temperature and physical parameters of the cloud. The NH$_3$ emission lines showed that spontaneous star formation is prevalent throughout the larger molecular cloud. The ionizing O6.5V star HD42088 (092.41489°, +20.48763°) is responsible for most of the excitation of the nebula (Jose et al., 2012; Chibueze et al., 2012; Jose, 2014), and contributes to the star formation in this region. The location of this star is seen in the Sh2-252 image in Fig. 4(A). Another O6V star, IRAS 06047+2040 located at $(\alpha, \delta)_{J2000} = (91.938° + 20.660°)$ is responsible for the ionization of the ultra-compact H II region called Sh2-252A. There is evidence for higher kinematic temperatures at Sh2-252A compared to the rest of the molecular cloud caused by the ionizing O6V star. This increase in temperature is an indication that the gas is hotter due to compression caused by an expanding H II region. Thus, a conclusion was drawn that although there is little evidence of large-scaled triggering, there is evidence of small-scaled triggering. They also investigated the distribution of YSOs and found 175 Class I, 268 Class II and 15 Transition Disks (Chibueze et al., 2012) in a 40′ × 40′ area.

### 2.1.8. Skull and Crossbones Nebula (NGC2467)

The Skull and Crossbones Nebula (NGC2467, Sh2-311, RCW16) is an H II region that forms part of the Puppis OB Association, located at a galactic anti-center distance of $5.0 \pm 0.4$ kpc within the Galaxy's third quadrant (Yadav et al., 2015). This region of the Milky Way has lower interstellar extinction compare to elsewhere, and enables us to examine star forming regions at greater distances towards the outer regions of the Galaxy (Yadav et al., 2015). NGC2467 has a spherical morphology and is ionized by a O6V star (∼5 Myr) (HD 64315) located at $(\alpha, \delta)_{J2000} = (118.085°, -26.430°)$ (Yadav et al., 2016). The distribution of YSOs along with the age of the O6V star and the age of the newer YSOs (∼1 Myr) suggests that trigger star formation is present especially near the exterior of the nebula (Yadav et al., 2015). A shock wave followed by an ionization front is the probable mechanism causing triggering. The rate of star formation within NGC2467 is $1.75 \times 10^{-3}$ M$_\odot$ yr$^{-1}$ and the triggered star formation rate is $0.42 - 0.90 \times 10^{-3}$ M$_\odot$ yr$^{-1}$, with an estimated 25%–50% of all stars are triggered (Snider et al., 2009). There is an open elongated cluster in this region called Haffner 18, as indicated by the square in NGC2467 image in Fig. 4(B). This cluster contains several late type O and B stars with largest star being a O6/7V star (Yadav et al., 2015). The distance to Haffner 18 was estimated to be $11.2 \pm 1.0$ kpc, and is located in the Norma-Cygnus Arm, one of the outer galactic arm. This shows that Haffner 18 is not part of the NGC2467 complex (Yadav et al., 2015).

### 2.1.9. Orion Nebula (M42)

The Orion Nebula (M42, NGC1976, Sh2-281, LBN974, Ced55d) is a well studied region of active high mass star formation at a distance of ∼410 pc to the Sun. This star-forming region is different to the rest of the regions within this study as the high mass stars are triggered through cloud-cloud collision (CCC). A supernova explosion may have initiated the expansion of the ONC, causing it to collide with the Orion A molecular cloud. During a CCC, clouds collide at supersonic speeds, gas is accumulated into dense cores. These cores eventually collapses and the formation of new stars is triggered (Kounkel et al., 2022; Chen et al., 2025). There is an age gradient in the Orion Nebula Complex (ONC), showing that a front of star formation is moving into the molecular cloud (Kounkel et al., 2022). This CCC took place ∼0.1 Myrs ago between the H<sub>II</sub> regions M42 and M43 (Fukui et al., 2018). Two molecular clouds, one blue-shifted and the other red-shifted (with high and low column densities respectively) collided forming O-type stars along the border of the blue-shifted cloud. The red-shifted cloud disintegrated as a result of the collision and subsequent ionization of the stars (Ohama et al., 2017). The relative velocity of the collision is ∼7 km s$^{-1}$ (Fukui et al., 2018). When examining the CO gas, the formation of high mass OB stars was triggered in localized regions as a result of the CCC (Fukui et al., 2018). These OB stars may have been triggered independently within the last Myrs (Fujita et al., 2021).

## 2.2. YSO classification

To determine and classify YSOs in each of these nine star-forming regions, a YSO classification scheme is applied. This YSO classification scheme is adapted from Koenig et al. (2012) and Chibueze et al. (2012). This leads to the classification of Class I, Class II and Transition Disks within each star-forming region. The results of this classification scheme are presented in Section 3.

### 2.2.1. Retrieving WISE data

Data is used from the Wide-field Infrared Survey Explorer mission (WISE) (Wright et al., 2010). WISE uses a scanning mode to make observations in four infrared bands (W1 (3.4 μm), W2 (4.6 μm), W3 (12 μm) and W4 (22 μm)). These bands are connected to 64 detectors to simultaneously map the entire sky, using a point source sensitivity of $5\sigma$ of ∼0.08, 0.1, 1, 6 mJy per band using a 40 cm diameter telescope (with 4 million pixels in total) (Wright et al., 2010). The site VizieR[1] was used to obtain the 2012 WISE data (Cutri et al., 2012) for each of the nine star-forming regions used in this study. This data includes the J, H and K band data and the respective band errors of the Two Micron All-Sky Survey (2MASS). A photometric uncertainty of < 0.2 mag limit to the W1, W2, W3 and W4 bands is applied. This WISE data will be used throughout to complete the classification of young stars for all nine regions.

### 2.2.2. Removing extragalactic objects

Any extragalactic objects are identified when the conditions in Eq. (1) is true. These extragalactic objects are removed according to the methods in Koenig et al. (2012). The color-color magnitudes for each of the 9 star-forming regions is presented in the Appendix Figs. A.9–A.11. These images shows that the dataset and the Extragalactic Objects are mixed among the YSOs that are to be classified. For this reason, the Extragalactic objects are removed to reduce inaccuracies in the classification. The corresponding color-magnitude diagrams are presented in the Appendix Figs. B.12–B.14.

$$\begin{aligned} W1 - W2 &< 0.46 \times (W2 - W3 - 1.7) \\ W1 - W2 &> -0.06 \times (W2 - W3 - 4.67) \\ W1 - W2 &< -1.0 \times (W2 - W3 - 5.1) \\ W1 - W2 &> 0.48 \times (W2 - W3 - 4.1) \\ W2 &> 12 \\ W2 - W3 &> 2.3 \end{aligned} \quad (1)$$

[1] https://vizier.cfa.harvard.edu

### 2.2.3. Removing AGNs

Any objects identified as active galactic nuclei is removed. These objects can often be misinterpreted as young stars due to emission at middle infrared wavelengths. To avoid misidentification, these objects are removed if either of the conditions in Eqs. (2) or (3) are met (Koenig et al., 2012).

$$\begin{aligned} W2 &> 1.9 \times (W2 - W3 + 3.16) \\ W2 &> -1.4 \times (W2 - W3 - 11.93) \\ W2 &> 13.5 \end{aligned} \quad (2)$$

$$\begin{aligned} W1 &> 1.9 \times (W2 - W3 + 2.55) \\ W1 &> 14 \end{aligned} \quad (3)$$

### 2.2.4. Removal of shock objects

There may be shock emission knots that contaminate the data set around the W2 (4.6 μm) band. These contaminants are removed if the conditions of Eq. (4) are met (Koenig et al., 2012).

$$\begin{aligned} W1 - W2 &> 1 \\ W2 - W3 &< 2 \end{aligned} \quad (4)$$

### 2.2.5. Removal of PAH objects

Another contaminant of the data is the emissions of polycyclic aromatic hydrocarbons (PAH) and can be seen as a point source in the W3 (12 μm) band. The PAH objects are identified and removed if either Eq. (5) or (6) are true (Koenig et al., 2012).

$$\begin{aligned} W1 - W2 &< 1 \\ W2 - W3 &> 4.9 \end{aligned} \quad (5)$$

$$\begin{aligned} W1 - W2 &< 0.25 \\ W2 - W3 &> 4.75 \end{aligned} \quad (6)$$

### 2.2.6. Classification of class I sources

The first YSOs in the data are identified as Class I sources. This classification seen in Eq. (7) is taken from Koenig et al. (2012).

$$\begin{aligned} W1 - W2 &> 1 \\ W2 - W3 &> 2 \end{aligned} \quad (7)$$

### 2.2.7. Classification of class II sources

Next, several Class II YSOs which have less red coloring, are identified with the conditions in Eqs. (8). Here the variable $\sigma$ refers to the standard deviation between the errors of bands W1 and W2, and W2 and W3, as seen in Eq. (9) (Koenig et al., 2012). Here, the errors are labeled as eW1, eW2, eW3 and eW4 for the W1, W2, W3 and W4 bands respectively as seen in Eq. (9).

$$\begin{aligned} W1 - W2 - \sigma_1 &> 0.25 \\ W2 - W3 - \sigma_2 &> 1 \end{aligned} \quad (8)$$

$$\begin{aligned} \sigma_1 &= \sigma(eW1 - eW2) \\ \sigma_2 &= \sigma(eW2 - eW3) \end{aligned} \quad (9)$$

### 2.2.8. Retrieving additional class II sources

Additional Class II YSOs are identified by examining the 2MASS K bands. This method is extracted from a combination of Koenig et al. (2012) and Chibueze et al. (2012) and is seen in Eq. (10). The variable $\sigma_3$ is the difference between the errors of the K band magnitudes and W1 magnitudes as seen in Eq. (11).

$$\begin{aligned} W1 - W2 - \sigma_1 &> 0.101 \\ K - W1 - \sigma_3 &> 0.0 \\ K - W1 - \sigma_3 &> -2.85714 \times (W1 - W2 - 0.101) + 1.05 \\ W1 &< 13.8 \end{aligned} \quad (10)$$

$$\sigma_3 = \sigma(eK - eW1) \quad (11)$$

**Table 1**
Table showing the estimated distance as well as the OB stars associated with each star-forming region. The first column represents the group the star-forming region is part of. The second column provides the name of the region. The third and fourth column provides galactic coordinates $(l, b)$ in decimal degrees. The fifth column includes the accepted $V_{LSR}$ as seen in Section 3. The sixth column shows the adopted parallax distance as mention for each region in Section 3. The seventh column indicates the Galactocentric distance for each region. The eighth column shows the names of the Ionizing stars in each region and their respective spectral types is presented in column nine, as well as their coordinates $(\alpha, \delta)_{J2000}$ in columns ten and eleven. The twelfth column provides the GAIA DR2 distances for each star.

| Group | Source name | $l$ | $b$ | $V_{LSR}$ (km s$^{-1}$) | Parallax distance (kpc) | Galactocentric distance (kpc) | Ionizing star | Spectral type | RA$_{(J2000)}$ | Dec$_{(J2000)}$ | GAIA DR2 Star distances (kpc) |
|---|---|---|---|---|---|---|---|---|---|---|---|
| Inner | SH2-19 | 0.770 | −1.240 | 2.0 | $1.37 \pm 0.3$ | 6.97 | HD 316398 | B8E | 267.297 | −28.898 | $1.572^{+0.261}_{-0.197}$ |
| | | | | | | | [GBB2012] E2 | B1V-B9V D | 267.573 | −28.894 | $4.962^{+3.136}_{-1.593}$ |
| | | | | | | | [GBB2012] E3 | B1V-B3V D | 267.571 | −28.895 | $2.721^{+0.448}_{-0.339}$ |
| | | | | | | | [GBB2012] E4 | B0V | 267.571 | −28.895 | $2.799^{+2.1}_{-0.891}$ |
| Inner | SH2-22 | 4.745 | −0.500 | 7.5 | $2.93 \pm 0.15$ | 5.43 | HD162978 | O8II((f)) C | 268.731 | −24.887 | $1.082^{+0.120}_{-0.098}$ |
| Inner | SH2-17 | 0.489 | −0.669 | −5.0 | $1.63^{+0.593}_{-0.303}$ | 6.71 | LHO 9 | O4-6If? C | 266.564 | −28.830 | $1.618^{+0.316}_{-0.229}$ |
| | | | | | | | LHO 44 | O7-9If? C | 266.562 | −28.828 | $1.645^{+1.276}_{-0.508}$ |
| | | | | | | | LHO 75 | WC9?d C | 266.559 | −28.827 | $1.711^{+0.145}_{-0.124}$ |
| | | | | | | | WR 102g | WC8 C | 266.563 | −28.828 | $1.546^{+0.633}_{-0.351}$ |
| Middle | M8 | 6.417 | −1.882 | 5.7 | $1.127^{+0.405}_{-0.176}$ | 7.22 | M8E-IR | | 271.222 | −24.445 | $0.947^{+0.787}_{-0.216}$ |
| | | | | | | | HD165052 | O6.5V((f))z C | 271.294 | −24.399 | $1.236^{+0.076}_{-0.068}$ |
| | | | | | | | 9 Sagittarii | O4V((f))z C | 270.969 | −24.361 | $1.156^{+0.149}_{-0.119}$ |
| | | | | | | | Herschel 36 | O8V+ C | 270.918 | −24.379 | $1.170^{+0.608}_{-0.301}$ |
| Middle | M16 | 17.300 | 0.197 | 14.8 | $1.717^{+0.258}_{-0.198}$ | 6.72 | HD168075 | O7V(n)((f))z | 274.650 | −13.793 | $1.961^{+0.354}_{-0.262}$ |
| | | | | | | | HD168137 | O8Vz C | 274.734 | −13.809 | $1.802^{+0.278}_{-0.214}$ |
| | | | | | | | NGC6611-222 | O7V((f))z C | 274.656 | −13.728 | $1.388^{+0.143}_{-0.119}$ |
| Middle | IC5070 | 84.954 | 0.103 | −0.4 | $0.793^{+0.046}_{-0.041}$ | 8.31 | HD 199579 | O6.5V((f))z C | 314.145 | 44.925 | $0.668^{+0.039}_{-0.035}$ |
| | | | | | | | 2MASSJ20555125+4352246 | O3.5III(f*)+O8 C | 313.964 | 43.873 | $0.918^{+0.054}_{-0.048}$ |
| Outer | NGC2467 | 243.525 | 0.696 | 51.7 | $4.16 \pm 0.38$ | 10.85 | HD64315 | O5.5Vz+O7V C | 118.085 | −26.430 | $7.508^{+2.557}_{-1.823}$ |
| Outer | SH2-252 | 190.272 | 1.004 | 8.0 | $2.10 \pm 0.03$ | 10.41 | HD42088 | O6V((f))z C | 92.415 | 20.488 | $1.664^{+0.189}_{-0.155}$ |
| | | | | | | | IRAS 06047+2040 | O6V | 91.938 | 20.660 | |
| Outer | M42 | 209.279 | −18.825 | 1.8 | $0.42 \pm 0.01$ | 8.69 | TCC51 | | 83.817 | −5.390 | $0.308^{+0.535}_{-0.040}$ |
| | | | | | | | HD 37041 | O9.5IVp C | 83.845 | −5.416 | $0.454^{+0.069}_{-0.053}$ |
| | | | | | | | HD 37061 | O9.5V C | 83.881 | −5.267 | $0.516^{+0.011}_{-0.011}$ |

### 2.2.9. Retrieving additional class I sources

Additional Class I sources were discovered from among the newly identified Class II sources from Section 2.2.8. The following condition in Eq. (12), identifies sources with very red color among these Class II YSOs and will reclassify them as Class I YSOs instead. This classification is taken from both Koenig et al. (2012) and Chibueze et al. (2012), using the K and W1 band magnitudes and $\sigma_3$ from Section 2.2.8.

$$K - W1 - \sigma_3 > -2.85714 \times (W1 - W2 - 0.401) + 1.9 \quad (12)$$

### 2.2.10. Classification of transition disks

Transition disks are defined as disks with less dust in their interior (van der Marel, 2023). These are disks around young stars in which planets may form (Espaillat et al., 2014). These disks show the evolutionary stage between optically thick disks around stars as in Class II objects and stellar objects that have dissolved disks as in Class III objects (Espaillat et al., 2014). To ensure that the classification of Transition Disks is accurate, the photometric error limit is less than 0.2 magnitudes for the W1 (3.4 µm), W2 (4.6 µm), and W4 (22 µm) bands. Next, the Transition Disks are identified when the following conditions of Eqs. (13) are met (Koenig et al., 2012).

$$W2 - W4 > 2.5$$
$$W1 < 14 \quad (13)$$

### 2.2.11. Removal of excess blue sources

Any Class II YSO sources with excess blue coloring is removed from the data set, as sources may be considered as T-Tauri stars when Eq. (14) is true (Koenig et al., 2012). T-Tauri stars are variable stars that have varying brightness (both periodically and randomly). These are low to intermediate mass stars ($< 8M_\odot$), and represent the intermediate evolutionary stage between protostars and low-mass main sequence stars (Herczeg and Hillenbrand, 2014).

$$W1 - W4 < 4 - 1.7 \times (W3 - W4) + 4.3 \quad (14)$$

### 2.2.12. Verification of YSO classification

To verify if the YSO classification was accurate, the results of the Monkey Head Nebula and NGC2467 were tested against previous studies. The results obtained by Chibueze et al. (2012) for the Monkey Head Nebula is: 175 Class I, 268 Class II and 15 Transition Disks. Since the number of YSO stars were similar to those in this study, this gave an indication that the classification may be correct. When analyzing NGC2467, the results were compared to Snider et al. (2009) who identified 14 YSOs as Class I/0, 16 as Class I/II and 12 as Class II. When comparing the YSOs found and classified by Snider et al. (2009), a ~70% accuracy rate was determined. This was estimated by matching the stars of NGC2467 by their coordinates and their class classification, to the results found in Snider et al. (2009) Table 1. The difference seen may be due to different areas investigated or the different surveys used.

## 3. Results

As seen in Fig. 1, the Galactic Plane is divided into three separate galactic longitudinal groups as the Inner, Middle and Outer Groups. Within each group, three star-forming regions have been identified with either possible or known signs of triggered star formation. These star-forming regions were selected as there is evidence of young stars within these regions as well as the presence of O/B type stars. These high mass stars drive expanding hydrogen ionization which trigger star formation as the ionization front moves over the cloud and compresses dense cores (Deharveng et al., 2005; Elmegreen, 2011). In the YSO distributions, ellipses were added to show possible region of triggered star formation. These possible regions were determined through the clustering of Class I sources (in yellow ellipse), surrounded by older Class II (in blue-dashed ellipses) and most of the Transition Disks were located outside of these ellipses. This onion-like layering will be noted if a high mass OB star is central to these stars forming (usually the OB star is within the molecular cloud) and the ionization front propagates outwards. If the OB star is outside the molecular cloud, then the newer stars will form along the ionization front and the older stars will be closer to the OB source.

The YSO distributions for each of these regions will be presented in this section. The background images are retrieved from the NASA/IPAC Infrared Science Archive using either Spitzer or WISE images for each source. In Figs. 2–4, areas where possible triggering occurs is indicated by yellow and blue-dashed ellipses. These ellipses are not the only finite regions of triggering within each star-forming region, but is a suggestion of where triggering may occur.

### 3.1. Inner group

The Inner Group is situated along the Galactic Plane at $355°-5°$ longitudes. This inner region is extremely active containing a supermassive black hole (Sgr A*) and is highly populated with HII regions. This region also contains the central molecule zone (CMZ), located within 500 pc of the Galactic Center (Walker et al., 2021; Morris and Serabyn, 1996). In the following subsections, the YSO distributions for Sh-22, Sh-19 and Sh-17 are presented.

#### 3.1.1. Sh2-22

Sh2-22 is located at $(l, b) = (4.745°, -0.500°)$. The background image Spitzer IRAC4 (8 μm) inverted heat image and the WISE data are centered on $(\alpha, \delta)_{J2000} = (268.834°, -25.106°)$ covering an area of $10' \times 10'$. The original WISE data contained 726 sources. After the YSO Classification a total of 129 YSOs were identified, of which there is 13 Class I, 59 Class II, 54 Transition Disks and 3 Unclassified sources. This YSO distribution for Sh2-22 is seen in Fig. 2(A). The ionizing Of-type star HD162978 is outside the image of Sh2-22 in Fig. 2(A). This star may be responsible for the triggering seen within the dense bright cores of the nebula. The Class I stars (gray squares) are located within the yellow ellipse, this is followed by the Class II stars (green circles) in the blue-dashed ellipse. The Transition Disks (blue triangles) are distributed throughout the rest of the molecular cloud, outside of these ellipses. The concentration of Class I stars shows localized triggering, and this may also indicate that there is a high-mass star hidden in this bright emission region. The triggering in this molecular cloud is localized to the bright emission regions where the ionization from HD162978 is dissipating the surrounding dense dust and revealing the young Class I stars. The radial velocity of $V_{LSR} = +7.5$ km s$^{-1}$ gives a distance of $2.7 \pm 1.6$ kpc (Lozinskaya et al., 1983). The kinematic distance is $2.93 \pm 0.15$ kpc (Reid et al., 2009), while GAIA DR2 gives $1.082^{+0.1196}_{-0.098}$ kpc for the ionizing star HD162978 (Bailer-Jones et al., 2018). We adopt $2.93 \pm 0.15$ kpc for the rest of this study (Reid et al., 2009).

#### 3.1.2. Sh2-19

Sh2-19 is located at $(l, b) = (0.770°, -1.240°)$. The Spitzer IRAC4 (8 μm) inverted heat image and WISE data are centered on $(\alpha, \delta)_{J2000} = (267.278°, -28.903°)$ with an area of $20' \times 20'$. The original dataset contained 2499 sources. After the YSO classification was applied, a total of 519 YSOs were found, of which are 35 Class I, 249 Class II, 207 Transition Disks and 28 Unclassified sources. The YSO distribution can be seen in Fig. 2(B). Possible triggered regions are denoted by the ellipses, indicating that there may be two triggered regions within this molecular cloud. The YSO distribution does indicate that this molecular cloud may be governed by spontaneous star formation, but has localized triggering. There are 4 ionizing stars detected near SH2-19 namely: HD316398, [GBB2012]E2, [GBB2012]E3, [GBB2012]E4. HD316398 is a B8E type star and is located in Fig. 2(B) and the remaining are outside this region. The location of HD316398 is $(\alpha, \delta)_{J2000} = (267.297°, -28.898°)$. The radial velocity of $H\alpha V_{LSR} = 2.0$ km s$^{-1}$ leading to a distance of $1.5 \pm 0.2$ kpc (Russeil, 2003). Using the $V_{LSR}$ the parallax distance is calculated as $1.37 \pm 0.30$ kpc (Reid et al., 2009). The average GAIA DR2 distance to the ionizing stars is $3.014^{+1.486}_{-0.755}$ kpc (Bailer-Jones et al., 2018). We adopt the distance $1.37 \pm 0.30$ kpc for the rest of this study.

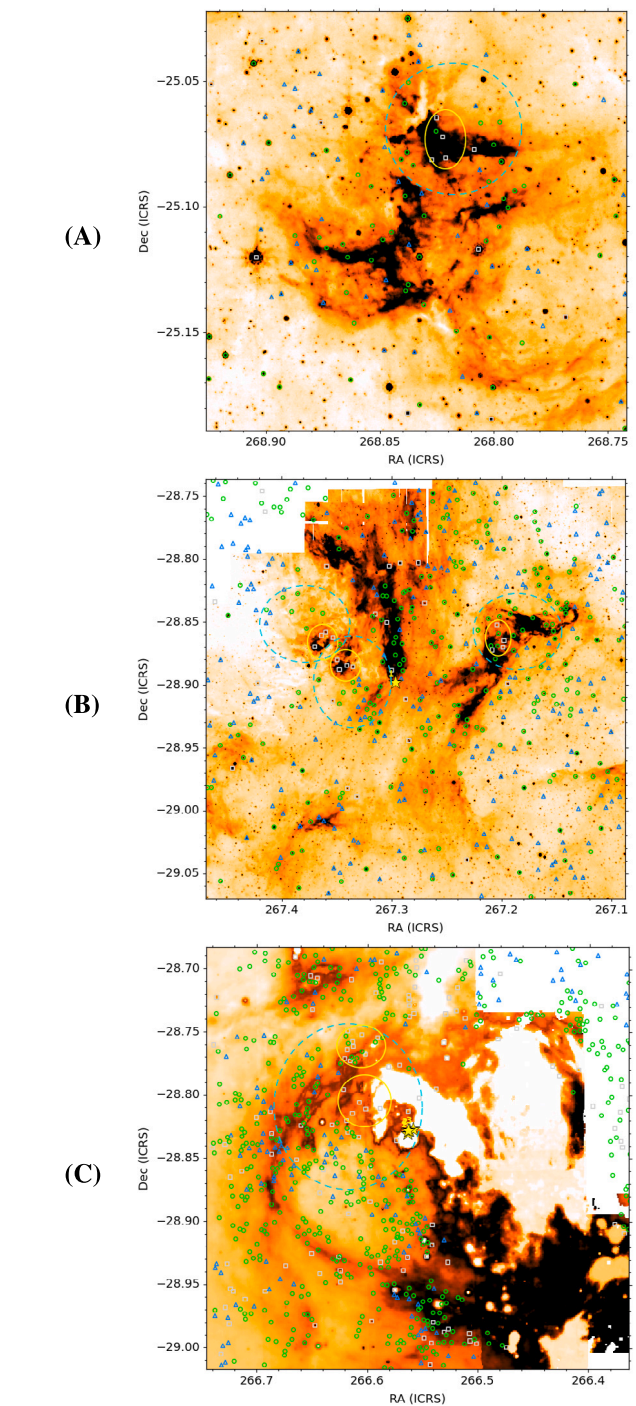

**Fig. 2.** The YSO distributions of the three star-forming regions of the Inner Group with Spitzer IRAC4 inverted heat background images; **(A)** is Sh2-22, **(B)** is Sh2-19, **(C)** is Sh2-17. The markers depict the locations of the Class I protostars (gray squares), Class II young stars (green circles) and Transition Disks (blue triangles) respectively. The yellow ellipses encapsulates the Class I stars, and the blue-dashed ellipses encompass Class II stars where possible triggering is seen. The locations of the ionizing stars are indicated by the yellow stars, in **(B)** 1: HD316398, **(C)** 1: LHO9, 2: LHO44, 3: LHO75, 4: WR102g.

#### 3.1.3. Sh2-17

Sh2-17 is located at $(l, b) = (0.489°, -0.669°)$. The Spitzer IRAC4 (8 μm) inverted heat image is used to present the HII region, as this wavelength showed the cloud structure the clearest. The image and WISE dataset is centered on $(\alpha, \delta)_{J2000} = (266.554°, -28.850°)$ encompassing an area of $20' \times 20'$. The original WISE dataset contained 2443 sources. Once the YSO classification was completed, a total of 715 YSOs were detected, of which 116 Class I, 459 Class II, 121 Transition Disks and 18 Unclassified sources were found. The YSO distribution of this H II region can be seen in Fig. 2(C). Several OB type stars (>50) were found to be clustered near the center of the image as indicated by the yellow stars in Fig. 2(C). Four O-type stars were selected to represent the larger OB cluster namely: LHO9 [O4-6If?C], LHO44 [O7-9If?C], WR102g [WC8C], LHO75 [WC9?dC]. The locations of LHO9 is $(\alpha, \delta)_{J2000} = (266.564°, -28.830°)$, LHO44 is $(\alpha, \delta)_{J2000} = (266.562°, -28.828°)$, WR102g is $(\alpha, \delta)_{J2000} = (266.563°, -28.828°)$, LHO75 is $(\alpha, \delta)_{J2000} = (266.559°, -28.827°)$. The regions in the yellow ellipses may show very early young stars forming, and could be an indication of localized triggered star formation. However, within the blue-ellipse both Class II and Transition Disks is detected and there is little clarity on an age gradient of younger (Class I) to older stars (Transition Disks). Therefore, Sh2-17 is a region mainly of spontaneous star formation, with new localized triggering taking place surrounding the OB stars. Using the $H\alpha V_{LSR} = -5.0$ km s$^{-1}$ gives a distance of $1.5 \pm 0.2$ kpc from Russeil (2003). The parallax distance is $2.75 \pm 0.28$ kpc (Reid et al., 2009). The average distance to the ionizing stars is $1.63^{+0.593}_{-0.303}$ kpc (Bailer-Jones et al., 2018), which will be adopted throughout this study.

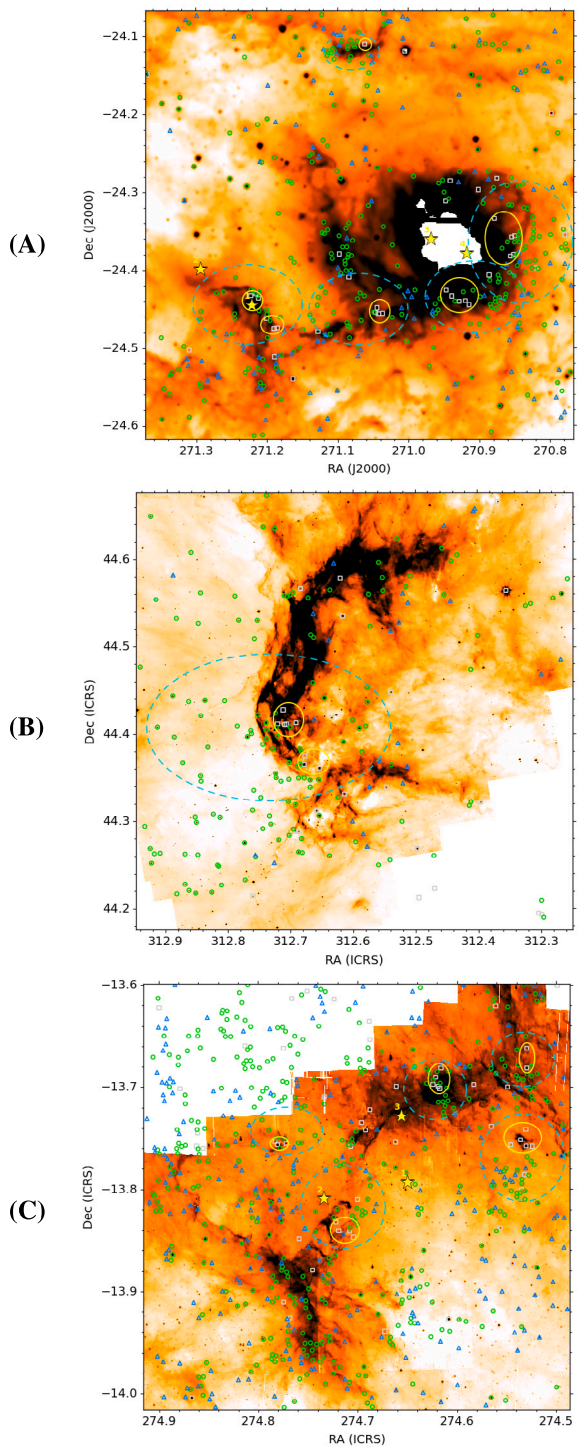

**Fig. 3.** The YSO distribution of the Middle Group H II regions: **(A)** is M8, **(B)** is IC5070 and **(C)** is M16. **(A)** has a WISE W3 and **(B, C)** has Spitzer IRAC4 inverted heat background images. The markers depict the locations of Class I protostars (gray squares), Class II stars (green circles) and Transition Disks (blue triangles) respectively. The yellow ellipses encapsulates the Class I stars, and the blue dashed ellipses encompass Class II stars where triggering is seen. The Transition Disks should mostly be outside of these circles. The location of ionizing stars are denoted by yellow stars; In (A), 1: M8E-IR, 2: HD165052, 3: 9 Sagittarii, 4: Herschel 36 and in **(C)**, 1: HD168075, 2: HD168137, 3: NGC6611-222.

### 3.2. Middle group

The Middle Group is allocated the galactic longitudes of 5° – 90° and 270°–355°, as seen in Fig. 1. This region is populated with star-forming regions that have been studied in depth throughout the years as they are often closer to the Solar Neighborhood compared to regions in either the Inner or Outer Group.

#### 3.2.1. Lagoon Nebula (M8)

The Lagoon Nebula is located at $(l, b) = (6.417°, -1.882°)$. The WISE 3 (12 μm) inverted heat images and WISE dataset is centered on $(\alpha, \delta)_{J2000} = (271.070°, -24.342°)$ covering an area of $33' \times 33'$. The original data set contained 6641 unidentified sources. After running the YSO classification, a total of 455 YSOs was detected. This includes 38 Class I, 279 Class II, 122 Transition Disks and 16 Unclassified sources. This YSO distribution can be seen in Fig. 3(A). The results show the evidence of localized triggering indicated by the yellow and blue ellipses. These triggered regions are driven by O-type stars such as HD165052, M8E-IR, 9 Sagittarii and Herschel 36, as denoted by yellow stars. This is evident by the Class I stars surrounding the bright region, followed by Class II and then Transition Disks. The most prominent triggered region is near the dense core driven by M8E-IR (Tothill et al., 2008). The Class I stars are located along the rim of the bright ridges of the cloud surrounded by older Class II stars, and Transition Disks located throughout the cloud. There is also evidence of small scale triggering near $(\alpha, \delta)_{J2000} = (271.04°, -24.1°)$. Using $H\alpha V_{LSR} = 5.7$ km s$^{-1}$ from Russeil (2003), the parallax distance is $1.54 \pm 0.10$ kpc (Reid et al., 2009). An accepted distance to M16 is ~1.25 kpc (Tiwari et al., 2020). The average distance of ionizing stars is $1.127^{+0.405}_{-0.176}$ kpc (Bailer-Jones et al., 2018) which will be adopted throughout this study.

#### 3.2.2. Pelican Nebula (IC 5070)

The Pelican Nebula (IC 5070) is located at $(l, b) = (84.954°, 0.103°)$. The Spitzer IRAC4 (8 μm) inverted heat image and the WISE dataset is centered on $(\alpha, \delta)_{J2000} = (312.598°, 44.427°)$ encompassing an area of $30' \times 30'$, resulting in 4537 sources. The YSO classification results in a total of 191 YSOs. This includes 27 Class I, 138 Class II, 25 Transition Disks and 1 Unclassified source, as seen in Fig. 3(B). Both ionizing stars 2MASSJ20555125+4352246 and HD199579 are located to the left outside the image. The yellow ellipse shows where Class I stars are located just inside the cloud and the blue ellipse encompasses a majority of the Class II stars beyond the bright ridge of the cloud. The Transition Disks are further into cloud. There is a clear indication of a large scale triggering in this nebula driven by the ionizing stars. This is evidenced by the Class II stars outside of the dense ridge, followed by the Class I stars within the dense regions of the cloud along the ridge. The older stars, the Transition Disks, are dispersed throughout the cloud. This is typical of the onion-like layering expected in triggered star formation. The common distance to IC5070 is 800 pc (Sicilia-Aguilar et al., 2024). However, when using H$\alpha$ $V_{LSR}$ = −0.4 km s$^{-1}$ (Russeil, 2003), the distance was calculated to be 1.50 ± 0.1 kpc (Reid et al., 2009). The average distance to ionizing stars is $0.793^{+0.0464}_{-0.0414}$ kpc (Bailer-Jones et al., 2018). The accepted distance to IC5070 is 0.8 kpc (Sicilia-Aguilar et al., 2024), therefore we adopt the distance of $0.793^{+0.0464}_{-0.0414}$ kpc throughout this paper.

Table 2
Table showing the distribution of YSOs for each of the nine H II regions. The first column represents which Inner, Middle or Outer Group this H II region is part of. The second column provides the name of the H II region. The third and fourth columns is the galactic coordinates. The fifth to ninth columns show the different YSO classes found for each H II region.

| Group | H II Region | $l$ | $b$ | Total YSOs | Class I | Class II | Transition disks | Unclassified sources |
|---|---|---|---|---|---|---|---|---|
| Inner | SH2-17 | 0.489° | −0.669° | 714 | 116 | 459 | 121 | 18 |
| Inner | SH2-19 | 0.770° | −1.240° | 519 | 35 | 249 | 207 | 28 |
| Inner | SH2-22 | 4.745° | −0.500° | 129 | 13 | 59 | 54 | 3 |
| Middle | M8 | 6.417° | −1.882° | 455 | 38 | 279 | 122 | 16 |
| Middle | M16 | 17.300° | 0.197° | 701 | 55 | 381 | 239 | 26 |
| Middle | IC5070 | 84.954° | 0.103° | 191 | 27 | 138 | 25 | 1 |
| Outer | SH2-252 | 190.272° | 1.004° | 287 | 117 | 157 | 13 | 0 |
| Outer | M42 | 209.279° | −18.825° | 470 | 171 | 293 | 5 | 1 |
| Outer | NGC2467 | 243.525° | 0.696° | 64 | 13 | 49 | 2 | 0 |

*3.2.3. Eagle Nebula (M16)*

The Eagle Nebula is located at $(l,b) = (17.300°, 0.197°)$. The background image used in Fig. 3(C) is the Spitzer IRAC4 (8 μm) inverted heat image. The image and the WISE dataset (3647 sources) is centered on $(\alpha,\delta)_{J2000} = (274.701°, -13.807°)$ covering an area of 25′ × 25′. The YSO classification identified a total of 701 YSOs. This contains 55 Class I, 381 Class II, 239 Transition Disks and 26 Unclassified sources. Although the YSO distribution displays an even distribution of young stars indicating spontaneity, however there is evidence of small localized triggering (as seen by the yellow and blue-dashed ellipses) near the pillar heads as the dust and gas dissipates with expanding radiation from the nearby ionizing O-type stars. These ionizing stars are HD168137 [O8Vz], NGC6611-222 [O7V((f))z] and HD16807 [O6.5V (f)]. Using the $V_{LSR}$ = 14.8 km s$^{-1}$ for a blue-shifted CO gas component from Nishimura et al. (2017). We calculated the distance to be 1.49 ± 0.08 kpc (Reid et al., 2009). The average distance to the ionizing stars is $1.717^{+0.258}_{-0.198}$ kpc (Bailer-Jones et al., 2018). The accepted distance to M16 is 1.8 kpc (Nishimura et al., 2017), therefore we adopt the distance of $1.717^{+0.258}_{-0.198}$ kpc throughout this paper.

*3.3. Outer group*

The Outer Group situated at 90°–270° galactic longitudes is located near the outskirts of the Galactic Plane. This region contains less activity and less dust extinction compared to the inner longitudes. Thus, it is easier to observe star formation at greater distances in a anti-galactic central direction (Yadav et al., 2015).

*3.3.1. Monkey Head Nebula (Sh2-252)*

The Monkey Head Nebula is located at $(l,b) = (190.272°, 1.004°)$. The Spitzer MIPS24 (24 μm) inverted heat image and the WISE dataset are centered at $(\alpha,\delta)_{J2000} = (92.274°, 20.55°)$ covering an area of 40′ × 40′ encompassing 5930 sources. The YSO classification found a total of 287 YSOs. This includes 117 Class I, 157 Class II, 13 Transition Disks and 0 Unclassified sources, as seen in Fig. 4(A). Although there is star formation throughout the molecular cloud, there is evidence of localized triggering driven by O-type stars HD42088 and IRAS 06047+2040 as described in Chibueze et al. (2012). The yellow stars indicate the locations of HD42088 and IRAS 06047+2040. Using a $V_{LSR}$ ~8.0 km s$^{-1}$ (Chibueze et al., 2012), the distance is calculated at 2.1 ± 0.03 kpc (Reid et al., 2009). The average distance to ionizing stars is $1.664^{+0.189}_{-0.155}$ kpc (Bailer-Jones et al., 2018). The accepted distance is 2.1 ± 0.03 kpc (Reid et al., 2009; Chibueze et al., 2012) and will be used throughout this paper.

*3.3.2. Skull and Crossbones Nebula (NGC2467)*

The Skull and Crossbones Nebula (NGC2467) is located at $(l,b) = (243.525°, 0.696°)$. The Spitzer IRAC4 (8 μm) inverted heat image and the WISE dataset are centered at $(\alpha,\delta)_{J2000} = (118.098°, -26.443°)$ covering an area of 20′×20′. The YSO classification found 64 YSOs. This includes 13 Class I, 49 Class II, 2 Transition Disks and 0 Unclassified sources, as seen in Fig. 4(B). There is evidence of triggering in this region confirming previous studies (Snider et al., 2009; Yadav et al., 2016). This triggering is indicated by the yellow and blue dashed ellipses in Fig. 4(B) caused by the O6V star HD 64315. The Class I stars are along the edge of the molecular ridge, followed by Class II stars. In Haffner 18 (white square), there is also evidence of triggered star formation driven by a large O6/7V star, and the younger stars are deep within the bright molecular core, surrounded by the older Class II stars. In Yadav et al. (2015), NGC2467 was calculated to be located in the Perseus Arm ( ~5.0 ± 0.4 kpc), however the Haffner 18 may be located further away at 11.2±1.0 kpc in the Outer Arm. Using a H$\alpha$ $V_{LSR}$ = 51.68 km s$^{-1}$ (Balser et al., 2011), the distance to NGC2467 is calculated at 4.16 ± 0.38 kpc (Reid et al., 2009). The GAIA DR2 distance to HD64315 is $7.508^{+2.557}_{-1.823}$ kpc (Bailer-Jones et al., 2018). The accepted to NGC2467 is 5.0 ± 0.4 kpc (Yadav et al., 2015). Therefore, we adopt the distance of 4.16 ± 0.38 kpc throughout this paper.

*3.3.3. Orion Nebula (M42)*

The Orion Nebula (M42) is located at $(l,b) = (209.279°, -18.825°)$. The Spitzer IRAC4 (8 μm) inverted heat image and the WISE dataset of 2527 unidentified sources is centered on $(\alpha,\delta)_{J2000} = (83.820°, -5.391°)$ encompassing an area of 30′ × 30′. Once the YSO classification was completed, a total of 470 YSOs were detected. This includes 171 Class I, 293 Class II and 5 Transition Disks and 1 Unclassified source, as seen in Fig. 4(C). The locations of O type stars TCC51, HD37041, HD37061 are indicated by the yellow stars. The location of TCC51 is $(\alpha,\delta)_{J2000} = (83.817°, -05.390°)$, HD37041 is $(\alpha,\delta)_{J2000} = (83.845°, -05.416°)$ and HD37061 is $(\alpha,\delta)_{J2000} = (083.881°, -05.267°)$. As mentioned in Section 2.1.9, a CCC may be responsible for the active star formation within this nebula. There is triggered star formation around the edges of the dense molecular cloud as the Class I stars are closest to the bright region, followed by Class II stars. This follows the pattern expected for triggered star forming regions. Using H$\alpha$ $V_{LSR}$ = 8.0 km s$^{-1}$ (Russeil, 2003), the distance calculated is 0.42 ± 0.01 kpc (Reid et al., 2009). The average GAIA DR2 distance to the ionizing stars is $0.426^{+0.205}_{-0.034}$ kpc. The accepted distance to M42 is ∼0.41 kpc (Ohama et al., 2017; Yamada et al., 2021). We adopt the distance 0.42 ± 0.01 kpc throughout this paper.

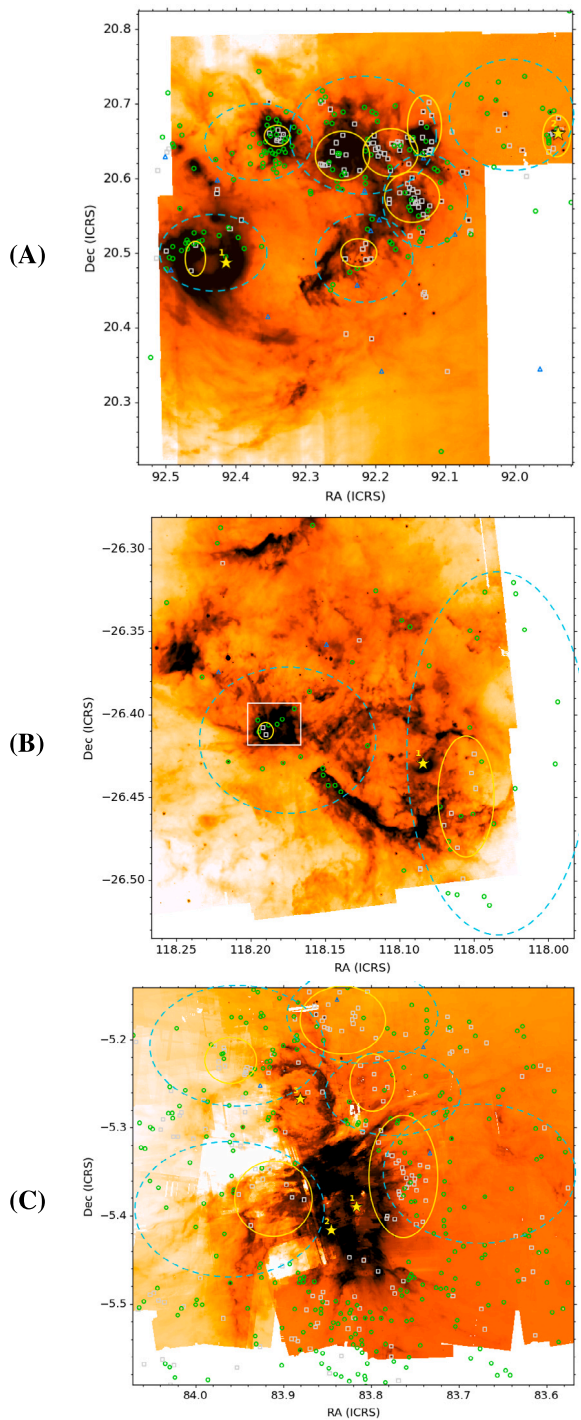

**Fig. 4.** The YSO distributions of three H<sub>II</sub> regions of the Outer Group. **(A)** is Sh-252 with Spitzer MIPS24 inverted heat image, **(B)** and **(C)** is NGC2467 and M42 respectively, both with Spitzer IRAC4 inverted heat images. The locations of the Class I protostars (gray squares), Class II young stars (green circles) and Transition Disks (blue triangles) and ionizing stars (yellow stares) is presented. In **(A)** 1: HD42088, 2: IRAS 06047+2040. In **(B)** 1: HD64315 and the square indicates the Haffner 18 open cluster. In **(C)** 1: TCC51, 2: HD37041, 3:HD 37061.

The results of the different YSO classes in each of the H<sub>II</sub> regions is tabulated in Table 2. The distances to the ionizing stars within each of the nine star-forming regions is retrieved from GAIA DR2[2] (Bailer-Jones et al., 2018). Table 1 presents distance information for each star-forming region and respective ionizing stars. The star-forming regions have varying distance when calculated with VLBI (Reid et al., 2009) compared to the GAIA DR2 distances (Bailer-Jones et al., 2018) of the ionizing stars. This may indicate a need for improved VLBI parallax measurements in the future.

## 4. Discussion

In this section the analysis of these results of the nine star-forming regions is presented. The first analysis is completed when comparing the YSO distributions of the star-forming regions across the different groups along the galactic longitudes and distance. The second analysis is of the YSO distributions throughout the spiral arms of the Milky Way Galaxy. This analysis is discussed in the following Sections 4.1 and 4.2. A few metrics are presented as well in Section 4.3.

### 4.1. Comparison across longitudes and distance

Inorder to compare and analyze the results of the star-forming regions throughout the Galaxy, YSO distributions are investigated along the absolute longitudes in Fig. 5 and the Galactocentric distance in Fig. 6. As a reminder the Inner Group is comprised of Sh2-17, Sh2-19 and Sh2-22; the Middle Group is comprised of M8, M16 and IC5070; and the Outer Group is comprised of Sh2-252, M42, NGC2467. The YSOs for each H<sub>II</sub> region is tabulated in Table 2.

In Fig. 5, the percentage of Class I YSOs is relatively low for the Inner (the average percentage of YSOs = 11.01 ± 4.81%) and Middle Group (average = 10.11 ± 3.49%) but this increases substantially in the Outer Group (average = 32.49 ± 10.77%). In Fig. 6, the Class I YSOs increases from <10% in the smaller distances (<6 kpc) to around ∼40% at the furthest distances (>10 kpc). A significant increase is seen ∼8–9 kpc which is at a similar distance to the Solar neighborhood ($R_0 = 8.24 ± 0.20$ kpc) (Bobylev, 2023). The slopes of the Class I trendline in Figs. 5 and 6 shows an increase from the Inner to Outer Group. These results show that local star-formation within the outer galactic longitudes and distances is higher than those closer to the Galactic Center.

In Fig. 5, the amount of Class II sources is larger than both the Transition Disks and the Class I sources across all nine H<sub>II</sub> regions, showing that there is ongoing star formation. The Class II trendline slope shows a slight increase from the Inner (average = 52.64±10.07%), Middle Group (average = 62.64 ± 9.02%) to Outer Group (average = 64.54 ± 11.09%). In Fig. 6, the percentage of Class II YSOs is generally much higher > 50% than the any of the other YSOs. There is an increase of Class II YSOs from ∼50% to 60 − 70% as distance increases from the galactic center. The reason for this increase towards the outer regions of the Galaxy is unknown and with more insights into these types of stars this increase may be clarified.

The Transition disks show an opposite trend in Figs. 5 and 6 to the Class I and Class II sources. In Fig. 5 the Transition Disks is substantially higher in the Inner Group (average = 32.93 ± 13.78%) and decreases to the Middle Group (average = 24.67 ± 10.67%) and the Outer Group (average = 2.91±1.74%). In Fig. 6, the percentage of Transition Disks has a significant decrease from the ∼6 kpc with ∼40% to 4% at the furthest distances from the galactic center. These results implies that there are fewer older stars within H<sub>II</sub> regions in the anti-galactic centric Outer group.

There is another class of sources called "Unclassified" as seen in Figs. 5. These YSOs have not been classified as either Class I, Class II

---

[2] https://dc.g-vo.org/gdr2dist/q/cone/form

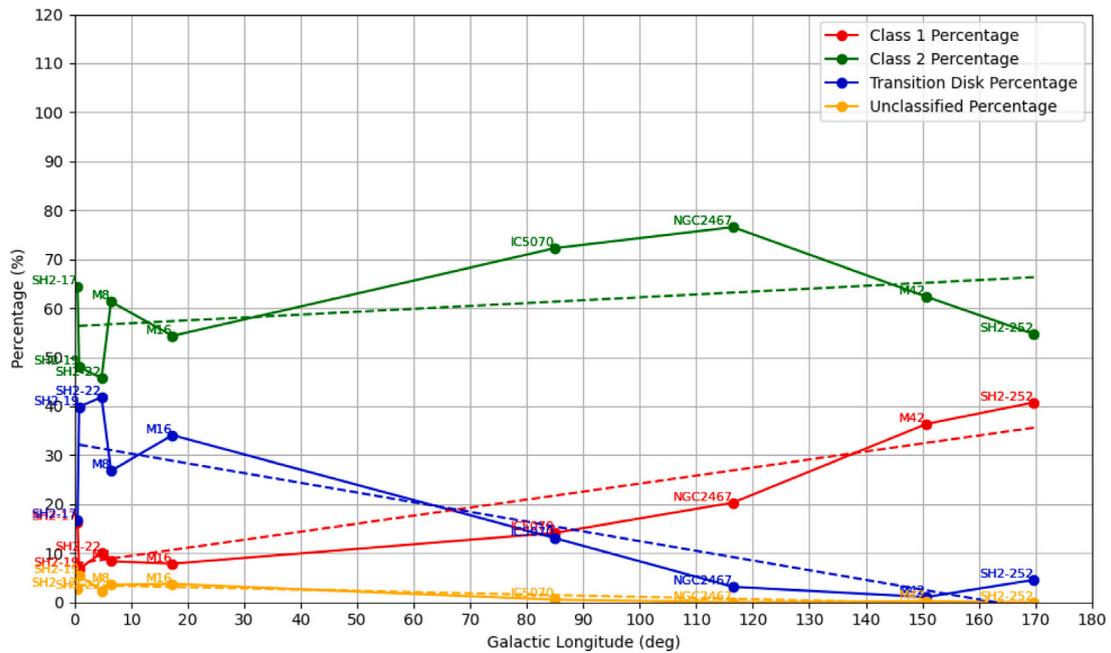

**Fig. 5.** The percentage of Class I, Class II, Transition Disks and Unclassified sources in each H$_{II}$ region with increasing absolute galactic longitudes. The solid lines show the actual percentage of YSOs and the dashed lines show the trendlines for each class.

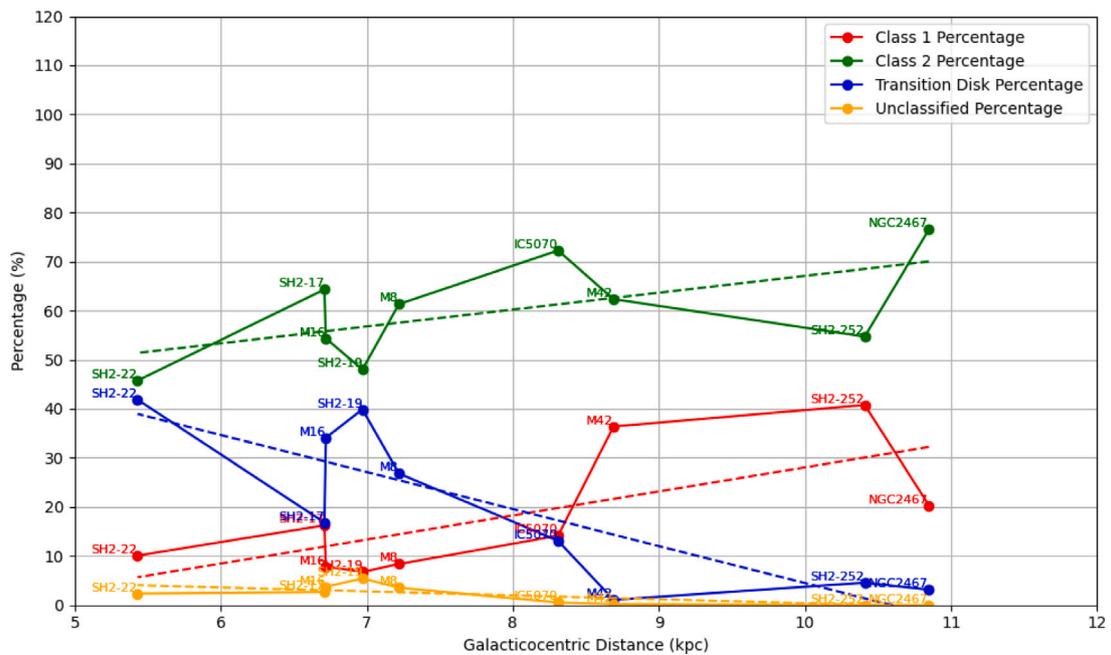

**Fig. 6.** The percentage of Class I, Class II, Transition Disks and Unclassified sources in each H$_{II}$ region with increasing Galactocentric distances. The solid lines show the actual percentage of YSOs and the dashed lines show the trendlines for each class.

or Transition Disk and may be another type of star which could possibly be a transitioning star between Class I and Class II phases of stars. As seen in Fig. 5, the amount of Unclassified YSOs remains relatively low throughout the Galaxy, but there are slightly more in the Inner Group (average = 3.41 ± 1.72%) than Middle Group (average = 2.58 ± 1.79%) and decreases in the Outer Group (average = 0.07 ± 0.12%). In Fig. 6 the amount of the 'Unclassified Sources' slightly decreases from ~5% to ~0%. If these 'Unclassified Sources' are transitioning stars, it may give an account for why the difference between Class II and Class I in the Inner Group is much larger than those seen in the Outer Group. These sources will need to be investigated in more detail to determine the kind of young stars they are as well as how fit into the evolutionary sequence of young stars.

These results show that there are more localized triggered star formation within the Outer Group than the Inner Group. This is supported by the trends of Class I and II objects. However, the Outer Group has significantly less Transition Disks, showing that there are fewer older stars at larger longitudes and distances. The difference of Class II and Class I is greater in the Inner Group as compared to the Outer Group. This shows there are more newer stars within the Outer Group but older stars locally within regions of the Inner Group. Our assumption of Unclassified YSOs being stars that transition between Class I and Class

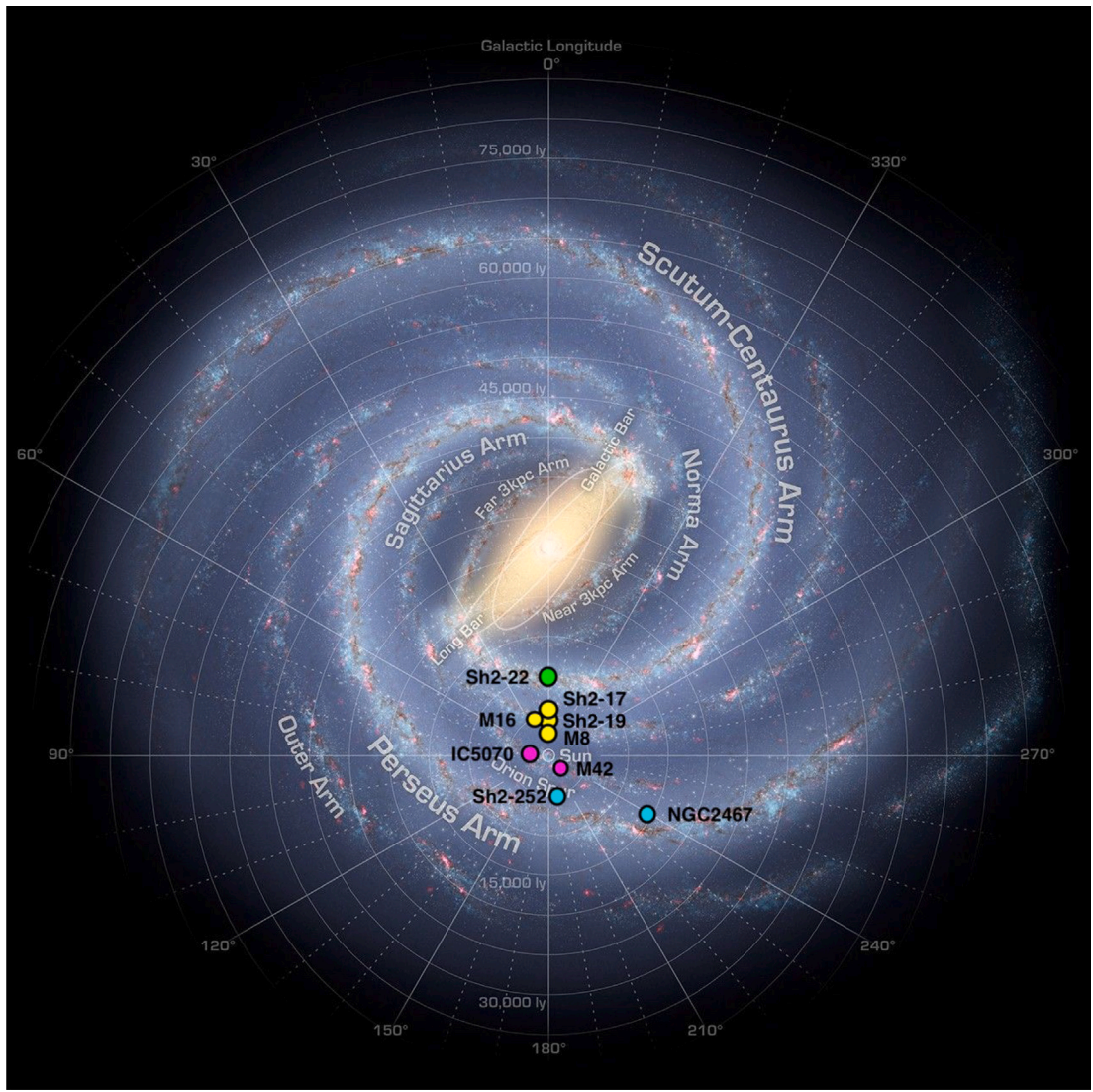

**Fig. 7.** An artists impression of the Milky Way Galaxy taken from NASA et al. (2013). The locations of the nine H<span style="font-variant:small-caps">ii</span> regions are indicated by the green dots in the Scutum-Centaurus Arm, yellow dots in Sagittarius Arm, yellow dots in Orion Spur, blue dots in the Perseus Arm.

II stars, may also indicate the presence of older stars within the Inner Group.

As this is a pilot study, it does not show an accurate representation of star-forming regions across the Galactic longitudes and distances, as this data only includes nine regions, spanning 5–11 kpc. To determine if these trends are true, more investigation of these star-forming regions is needed, along with a larger dataset across all longitudes and distances.

*4.2. Comparison across galactic arms*

The nine star-forming regions can be distinguished by their locations within the spiral arms as seen in Fig. 7. To study how the localized triggered star formation in these regions changes across the different galactic arms, will assist with understanding how the larger galactic environment may influence localized star formation within molecular clouds. The innermost spiral arm is Scutum Centaurus, followed by the Sagittarius Arm, Orion Spur, and the Perseus Arm is the furthest from the Galactic Center.

The average percentage of YSOs in each of spiral arm is tabulated in Table 3. In Table 3, the average number of Class I YSOs generally increases from the inner spiral arm, Scutum-Centaurus, to the outer arm, Perseus. The clear increase between the Sagittarius and Orion arms shows the possibility of newer stars being triggered in further from the Galactic Center. This could be an indication that these star-forming regions in the outer spiral regions may be younger and more active.

In Table 3, the average number of Class II YSOs is higher than the rest of the sources with a slight increase from the inner to outer spiral arms. This shows that there is generally more older stars in all the regions of the Galaxy, however there may be more Class II stars in the outer regions of the Galaxy. This is very similar to the pattern seen of average Class II YSOs in Section 4.1 and in Fig. 5.

In Table 3, the average number of Transition Disks decreases sharply between Sagittarius and the Orion Spur. This is may be an indication that there are fewer older stars in the furthest spiral arms, and may indicate a difference between triggered star formation within the inner spiral arms compared to the outer spiral arms. This pattern will need to be verified by analyzing star formation in more regions across all the spiral arms.

In Table 3, the average number of Unclassified sources is slightly higher in the Scutum Centaurus Arm and decreases to 0.00% in the Perseus Arm. Our hypothesis that these unclassified stars may be transitioning between Class I and Class II stars, can indicate that there are more stars situated between a Class I and Class II phase in the Scutum and Sagittarius arms as the difference between Class II and Class I stars is larger in these spiral arms.

Table 3
Table showing the average of different YSOs in the four spiral arms. The first column indicates the spiral arms. The second column indicates which HII regions are located in each spiral arm. The third to sixth columns indicate the average number of stars of each region within the spiral arms.

| Spiral arm | Star-forming regions | Average Class1 percentage | Average Class2 percentage | Average transition disks percentage | Average unclassified percentage |
|---|---|---|---|---|---|
| Scutum Centaurus | Sh2-22 | 10.078 | 45.736 | 41.860 | 2.326 |
| Sagittarius-Carina | Sh2-19, Sh2-17, M8, M16 | 10.209 | 57.238 | 28.828 | 3.682 |
| Orion Spur | IC5070, M42 | 29.955 | 65.204 | 4.539 | 0.303 |
| Perseus | NGC2467, SH2-252 | 37.037 | 58.689 | 4.274 | 0.000 |

## 4.3. Metrics of YSO distributions

The stellar evolution of the YSOs across the nine star-forming regions can be analyzed by looking at the ratios of Class I, Class II and Transition Disks. In Fig. 8(A), the ratio $\frac{N_{II}}{N_I}$ shows whether the evolutionary stage of when stars transition from Class I to Class II compared to the distance from the galactic center. Similarly, in Fig. 8(B), the ratio of $\frac{N_{TD}}{(N_{II}+N_{TD})}$ shows if the ratio of older transition disks over all evolved YSOs is proportional to the distance from the galactic center. If a trend appears in these images then evolution of these YSOs may be governed by the galactic structure, however if no trend is seen, then the evolution may be dominated by the conditions of the local molecular cloud. In Fig. 8(A), a trend is not clearly seen among the star-forming regions with increasing distance. Thus, the Class I and Class II YSO evolution may be governed by the local molecular cloud environment rather than the galactic environment. In Fig. 8(B), a slight decreasing trend may indicate that the star-forming regions may be governed by the galactic structure.

The combination of the ratios $\frac{N_{TD}}{(N_{II}+N_{TD})}$ and $\frac{N_{II}}{N_I}$ indicates the evolution of YSOs within all star-forming regions as seen in Fig. 8(C). The $\frac{N_{II}}{N_I}$ values are proportional with the age of the stellar clusters, and $\frac{N_{TD}}{(N_{II}+N_{TD})}$ measures the disk evolutionary stage. These two metrics together indicate if older star forming environments have a higher fraction of transition disks. An upward trend may support the idea that the stellar disks are cleared more as the clusters age. However if there is no trend, the stellar disks are cleared not just due to the age of the star cluster/environment but is also influenced by local conditions of the star-forming region. In Fig. 8(C) a slight increase is seen indicating that the galactic environment may influence the evolution of YSOs from Class I to Transition Disks. This is supported when noticing that the star-forming regions further away (Sh2-252, M42 and NGC2467) have the lowest values compared to those closer to the galactic center.

In this pilot study, the ratios of $\frac{N_{II}}{N_I}$ and $\frac{N_{TD}}{(N_{II}+N_{TD})}$ are used to indicate evolutionary changes of the YSOs throughout the star forming regions. However, since there is not enough data the trend is not clearly visible, no conclusion can be drawn regarding if the stellar evolution within the star-forming regions is dominated by either galactic structure or the local cloud environment. Additional metrics such as spatial distance from OB stars may provide insights into understanding how mixed YSOs are and when OB stars drive star-formation. An example of an additional metric is the Gini parameter which can determine how evenly distributed the different classes of stars are within each region. Using similar metrics with larger datasets may enhance these results in future studies.

## 4.4. Overall implications

The Class I YSOs increases from the Inner to the Outer Group and the Scutum-Centaurus to the Perseus spiral arms. The amount of Class II sources is relatively high, but slightly elevated near the outer regions of the Galaxy. Both these trends of Class I and II shows that there is evidence of heightened local star formation taking place in the Outer Group and the Orion Spur and Perseus spiral arms. There is also far fewer Transition Disks within this region giving an indication that there are fewer old stars - showing star formation is still active in these regions.

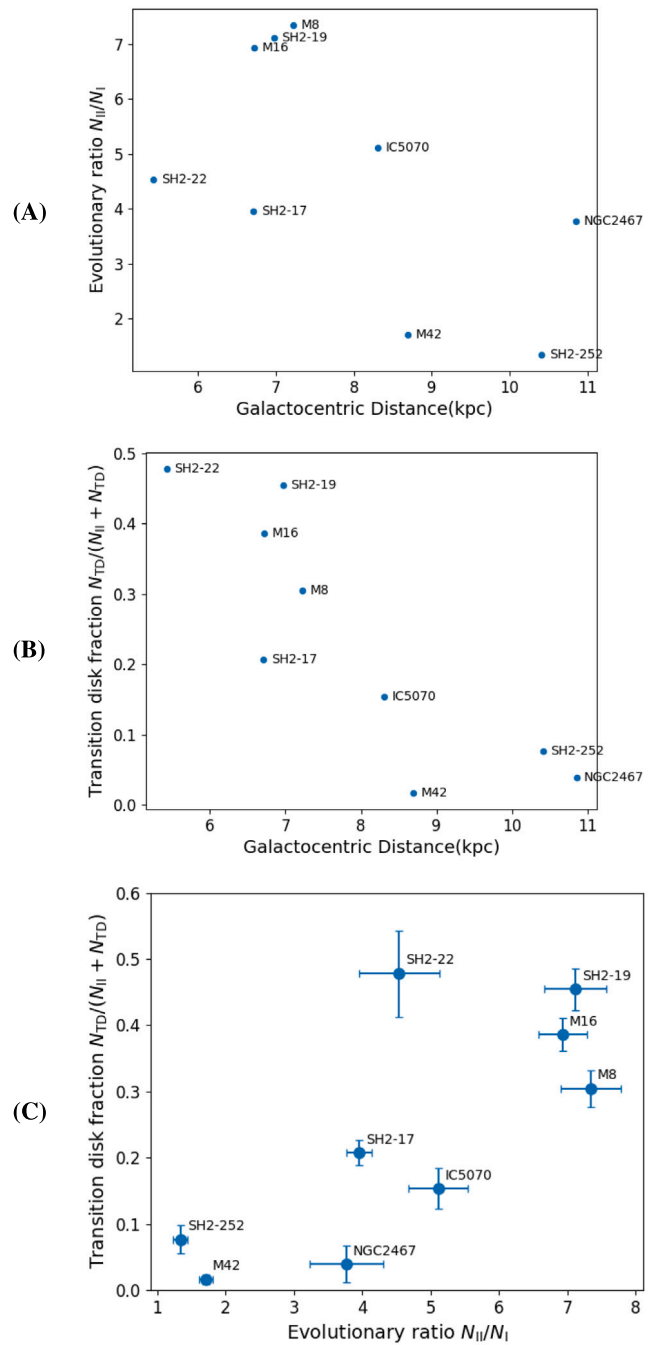

**Fig. 8.** Ratios of **(A)** $\frac{N_{II}}{N_I}$ and **(B)** $\frac{N_{TD}}{(N_{II}+N_{TD})}$ compared to the Galactocentric distances for each of the nine star-forming regions. In **(C)**, these ratios are compared to one another representing the evolution of Transition Disks compared to Class II and Class I.

These results may be supported by a study done by Enokiya and Fukui (2021), where the amount of diffuse gas in the inner regions (3 - 4 kpc) of the Galaxy is ∼10%–20%, and at larger distances (∼15 kpc) is 50%. This increase correlates with a decreasing surface density (Roman-Duval et al., 2016). This increase in gas may give rise to younger stars being formed through the triggering process in the outer regions of the Galaxy. The results within this study support previous literature stating that the star formation rate near the Galactic Center is less than 1–2 orders of magnitude than the expected theoretical rate. The star formation in the Galactic Center is suppressed and fewer stars are formed, although there is enough material and molecular clouds for higher star formation rates (Enokiya and Fukui, 2021).

When studying the formation of the Milky Way Galaxy Bar, it was noted that stars formed within the inner disk regions before the bar was formed has drifted to the outer disk regions (Baba, 2025). This migration is driven by "bar quenching" and the development of strong spiral arms. These spiral arms increases the star formation within the inner galactic disk and supports the transport of angular momentum to the outer disk region. Therefore, there is an increase in new stars during the bar formation epoch towards the outer disk regions. This also means that these stars have similar metallic properties to those within the inner regions (Baba, 2025). The results showed that there is peaks of star formation at galactic radii of $2 < R < 3$ kpc, ∼8 kpc (Solar neighborhood) and $11 < R < 12$ kpc, around the time of the bar formation (Baba, 2025). Understanding how the bar was formed and how it influences the metallicity of the galaxy, may shed light on the trends seen in this pilot study.

It is important to note that these results present may show the local environment within each individual star-forming region instead of a general representation of how the location within the Galaxy influences local star formation. This suggestion is drawn as the number of regions within this study is not a real representation of the star-forming regions throughout the entire Galaxy. The decrease in the number of Transition Disks and increase in Class I and II stars in the outer regions of the Galaxy shows that there may be more localized triggered star formation in outer regions. Thus, it is a possibility that the location of the triggered region within the Galaxy may influence the local YSO star formation within the molecular cloud. In Fig. 7, the HII regions are situated in a near straight line from the Galactic Center to the Sun, except for NGC2467. Thus, our results may be biased towards a very narrow region in the Galaxy and is not an accurate representation of star-forming regions throughout each spiral arm throughout the Galaxy. These results form the foundation for future investigations based on this pilot study.

## 5. Conclusions

This paper presents the pilot study that investigates how YSO distributions changes throughout the Galaxy. The aim was to determine if local triggered star formation within a star-forming region changes with respect to the location within the Galaxy. This was completed by studying the YSO distributions within nine star-forming regions located at various galactic longitudes. These galactic longitudes were grouped separately into Inner, Middle and Outer Groups as seen in Fig. 1.

This study provides a proof of concept that YSO distributions within triggered regions changes in relation to their location within the Galaxy. The number of Class I, Class II, Transition Disks and Unclassified sources were analyzed and presented in Figs. 2–4. The trends of these YSO distributions were analyzed in Figures 5 and 6. These Figures shows that there is a significant decline in the amount of Transition Disks with increasing longitudes and Galactocentric distance. This reveals that there may be fewer older stars in the outer triggered star-forming regions than those closer to the Galactic Center. The increase of Class I and Class II stars in the outer regions of the Galaxy also reveal that there may be local environmental stability encouraging the growth of new stars in molecular clouds, whereas the inner part of the Galaxy may be more unstable due to the turbulent and violent environment of the central molecular zone and the supermassive blackhole Sgr A*. This instability may cause disruption leading to the formation of fewer young Class I stars and retaining of more older Class II and Transition Disks.

The ratios of numbers of Class I, Class II and Transition Disks compared to the Galactocentric distances were analyzed in Fig. 8. In Fig. 8(A), no definitive trend is seen to determine if the star-formation is governed by the galactic environment and it is suggested this may be governed by the local environment. In Fig. 8(B), slight downward trend was seen indicating that the evolution of Transition Disks may be governed by the galactic environment. In Fig. 8(C), a slight increase is seen indicating that the galactic environment may have a role in number of Transition Disk. This is supported when the star-forming regions furthest from the galactic center has the lowest $\frac{N_{TD}}{(N_{II}+N_{TD})}$ values.

As this is a pilot study the results are not conclusive due to the small sample size of star-forming regions. The trends and ratios of the classes of stars as well as the spatial distributions of the YSOs to the OB stars provide interesting insights which can be explored with subsequent studies. Future studies may compare these results to hydrodynamical simulations of stellar, molecular or galaxy evolution. Some examples of models which could be used is the ZV14 model which investigates GMC formation (Vázquez-Semadeni et al., 2018), numerical simulations of gravoturbulent fragmentation using the ratio of different classes of stars (Schmeja et al., 2005), and hydrodynamical simulations used to predict various wavelength fluxes for periodic accretion of YSOs (MacFarlane et al., 2019). Using the YSO distributions and spatial distance of YSOs for a larger sample size, along with simulations will provide a clearer understanding of how stars form within their local molecular cloud environments and how this changes throughout the galaxy. This will also give insight into how the galactic environment influences local star formation.

## CRediT authorship contribution statement

**Annarien G. Headley:** Writing – original draft, Visualization, Software, Methodology, Investigation, Formal analysis, Data curation, Conceptualization. **James O. Chibueze:** Writing – review & editing, Validation, Supervision, Conceptualization.

## Declaration of competing interest

The authors declare that they have no known competing financial interests or personal relationships that could have appeared to influence the work reported in this paper.

## Acknowledgments

We would like to acknowledge the support the University of South Africa, South Africa for providing financial and academic support towards this project. We acknowegde the support from the Italian Ministry of Foreign Affairs and International Cooperation (MAECI Grant Number ZA18GR02) and the South African Department of Science and Technology's National Research Foundation (DST-NRF Grant Number 113121) as part of the ISARP RADIOSKY2020 Joint Research Scheme . The research has made use of the SIMBAD database, operated at CDS (Strasbourg, France), as well as NASA's Astrophysics Data System Bibliographic Services. This research has made use of the VizieR catalog access tool, CDS, Strasbourg Astronomical Observatory, France (DOI : 10.26093/cds/vizier). AllWISE makes use of data from WISE, which is a joint project of the University of California, Los Angeles, and the Jet Propulsion Laboratory/California Institute of Technology, and NEOWISE, which is a project of the Jet Propulsion Laboratory/California Institute of Technology. WISE and NEOWISE are funded by the National Aeronautics and Space Administration. This work is based [in part] on observations made with the Spitzer Space Telescope, which was operated by the Jet Propulsion Laboratory, California Institute of Technology under a contract with NASA. The distance data from GAIA DR2 is retrieved from provided by the German Astrophysical Virtual Observatory (GAVO) data center.

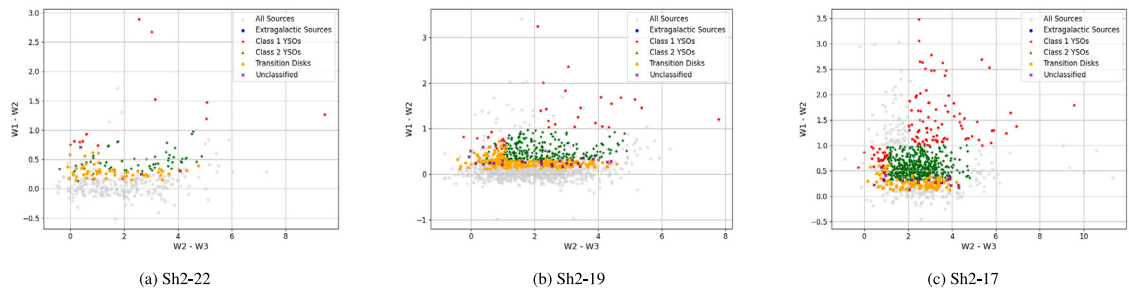

Fig. A.9. Inner group color-color magnitudes.

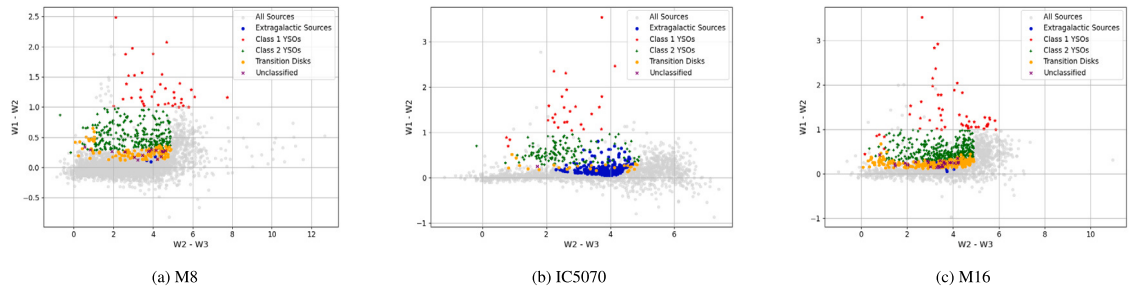

Fig. A.10. Middle group color-color magnitudes.

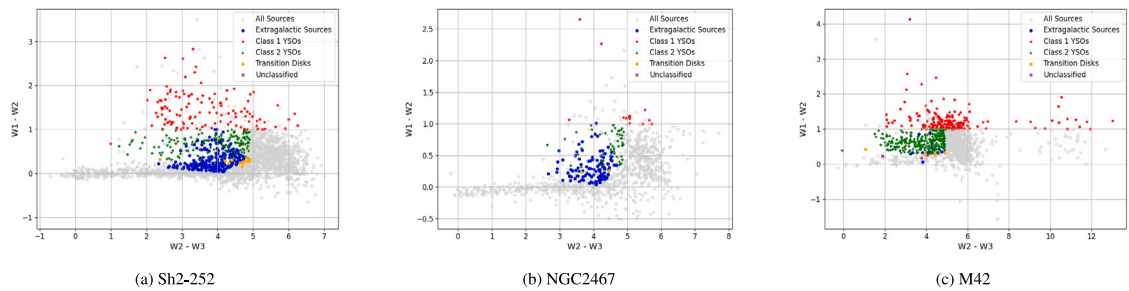

Fig. A.11. Outer group color-color magnitudes.

**Appendix A. Color-color magnitude**

In this section the color-color magnitude images are presented for each of the star-forming regions separated into the Inner, Middle and Outer groups. These $w_1 - w_2$ versus $w_2 - w_3$ graphs shows all sources of the respective WISE datasets in gray, extragalactic sources in blue, Class I YSOs in red, Class II YSOs in green, Transition Disks in yellow and any Unclassified sources in purple.

**Appendix B. Color magnitude diagrams**

In this section the color-magnitude diagrams are presented for each of the star-forming regions separated into the Inner, Middle and Outer groups. These $J$ versus $J - H$ magnitude graphs shows all sources of the respective WISE datasets in gray, extragalactic sources in blue, Class I YSOs in red, Class II YSOs in green, Transition Disks in yellow and any Unclassified sources in purple.

**Appendix C. Supplementary data**

Supplementary material related to this article can be found online at https://doi.org/10.1016/j.newast.2025.102482.

This is the Supplementary data for this study in the form of csv files. This includes the results of the YSOs for each of the 9 star forming regions. Column 1 assigns a unique identified number for each YSO. Columns 2 and 3 indicate the coordinates of RA and Declination in degrees (J2000). Column 4 indicates the YSO Class (either 1 = Class I, 2 = Class II, TD = Transition Disks or U = Unclassified. Columns 5-12 provide the WISE band magnitudes and the respective errors. Columns 13-18 provide the 2MASS Band magnitudes and the respective errors.

**Data availability**

The data used in this pilot study can be retrieved from the Vizier (https://vizier.cfa.harvard.edu), and NASA/IPAC Infrared Science Archive (https://irsa.ipac.caltech.edu). The distances were obtained from the GAIA DR2 from German Astrophysical Virtual Observatory (GAVO) (https://dc.g-vo.org/gdr2dist/q/cone/form) data center as well as The Bar And Spiral Structure Legacy (BeSSeL) Survey (http://bessel.vlbi-astrometry.org/home). The csv files containing the identified YSOs within each of the nine star forming regions is available in the Supplematary data section.

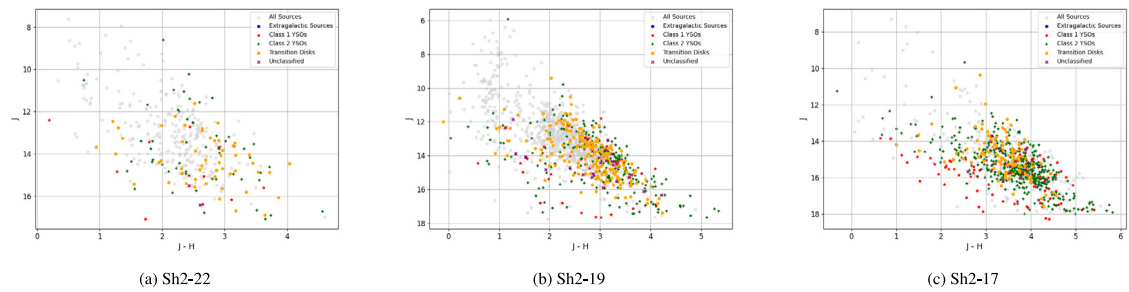

Fig. B.12. Inner group color magnitude diagrams.

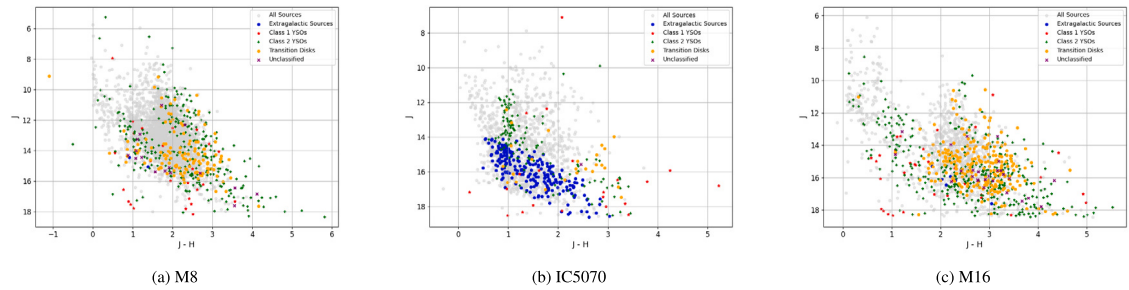

Fig. B.13. Middle group color magnitude diagrams.

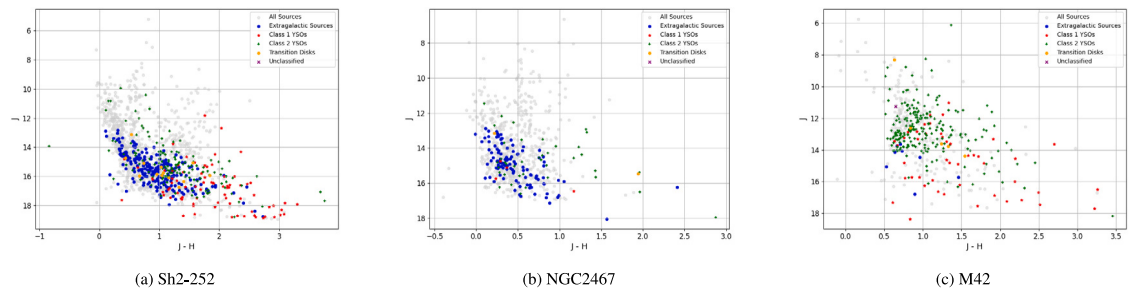

Fig. B.14. Outer group color magnitude diagrams.

## References

Avedisova, V.S., Palous, J., 1989. Kinematics of Star Forming Regions. Bull. Astron. Inst. Czech. 40, 42.

Baba, J., 2025. Influence of Bar Formation on Star Formation Segregation and Stellar Migration: Implications for Variations in the Age Distribution of Milky Way Disk Stars. http://dx.doi.org/10.48550/arXiv.2505.16528, ArXiv E-Prints. arXiv:2505.16528.

Bailer-Jones, C., Rybizki, J., Fouesneau, M., Mantelet, G., Andrae, R., 2018. Estimated distances to 1.33 billion stars in gaia DR2. VO resource provided by the GAVO Data Center. URL https://dc.g-vo.org/gdr2dist/q/cone/info.

Balser, D.S., Rood, R.T., Bania, T.M., Anderson, L.D., 2011. H II Region Metallicity Distribution in the Milky Way Disk. ApJ 738 (1), 27. http://dx.doi.org/10.1088/0004-637X/738/1/27, arXiv:1106.1660.

Beltrán, M.T., Massi, F., López, R., Girart, J.M., Estalella, R., 2009. The stellar population and complex structure of the bright-rimmed cloud IC 1396N. A&A 504 (1), 97–107. http://dx.doi.org/10.1051/0004-6361/200811540, arXiv:0902.4543.

Bieging, J.H., Peters, W.L., Vila Vilaro, B., Schlottman, K., Kulesa, C., 2009. Sequential Star Formation in the Sh 254-258 Molecular Cloud: Heinrich Hertz Telescope Maps of CO J=2-1 and 3-2 Emission. AJ 138 (3), 975–985. http://dx.doi.org/10.1088/0004-6256/138/3/975.

Blaauw, A., 1964. The O Associations in the Solar Neighborhood. ARAA 2, 213. http://dx.doi.org/10.1146/annurev.aa.02.090164.001241.

Bobylev, V.V., 2023. Estimation of the Galactocentric Distance of the Sun from Cepheids Close to the Solar Circle. Astron. Lett. 49 (9), 493–500. http://dx.doi.org/10.1134/S1063773723090025, arXiv:2312.05500.

Carpenter, J.M., Snell, R.L., Schloerb, F.P., 1995. Star Formation in the Gemini OB1 Molecular Cloud Complex. ApJ 450, 201. http://dx.doi.org/10.1086/176132.

Chen, E., Chen, X., Fang, M., Chen, X., He, Q., Yang, T., 2025. Bullet Shooting Cloud–Cloud Collision in MIR Bubble N65. ApJ 985 (1), 120. http://dx.doi.org/10.3847/1538-4357/adc9a3, arXiv:2505.10970.

Chen, W.P., Lee, H.T., Sanchawala, K., 2007. Triggered star formation in OB associations. In: Elmegreen, B.G., Palous, J. (Eds.), Triggered Star Formation in a Turbulent ISM. In: IAU Symposium, vol. 237, pp. 278–282. http://dx.doi.org/10.1017/S1743921307001603, arXiv:astro-ph/0611529.

Chibueze, J.O., Imura, K., Omodaka, T., Handa, T., Nagayama, T., Fujisawa, K., Sunada, K., Nakano, M., Kamezaki, T., Yamaguchi, Y., Sekido, M., 2012. STAR FORMATION IN THE MOLECULAR CLOUD associated WITH THE MONKEY head NEBULA: Sequential OR spontaneoUS? Astrophys. J. 762 (1), 17. http://dx.doi.org/10.1088/0004-637X/762/1/17.

Choudhury, R., Mookerjea, B., Bhatt, H.C., 2010. TRIGGERED STAR FORMATION AND YOUNG STELLAR population IN BRIGHT-rimmed CLOUD SFO 38. Astrophys. J. 717 (2), 1067. http://dx.doi.org/10.1088/0004-637X/717/2/1067.

Cutri, R.M., Wright, E.L., Conrow, T., Bauer, J., Benford, D., Brandenburg, H., Dailey, J., Eisenhardt, P.R.M., Evans, T., Fajardo-Acosta, S., Fowler, J., Gelino, C., Grillmair, C., Harbut, M., Hoffman, D., Jarrett, T., Kirkpatrick, J.D., Leisawitz, D., Liu, W., Mainzer, A., Marsh, K., Masci, F., McCallon, H., Padgett, D., Ressler, M.E., Royer, D., Skrutskie, M.F., Stanford, S.A., Wyatt, P.L., Tholen, D., Tsai, C.W., Wachter, S., Wheelock, S.L., Yan, L., Alles, R., Beck, R., Grav, T., Masiero, J., McCollum, B., McGehee, P., Papin, M., Wittman, M., 2012. Explanatory Supplement to the WISE All-Sky Data Release Products. p. 1, Explanatory Supplement to the WISE All-Sky Data Release Products. https://wise2.ipac.caltech.edu/docs/release/allsky/expsup/sec2_2a.html.

Damiani, F., Prisinzano, L., Micela, G., Sciortino, S., 2019. Wide-area photometric and astrometric (Gaia DR2) study of the young cluster NGC 6530. A&A 623, A25. http://dx.doi.org/10.1051/0004-6361/201833877, arXiv:1812.11402.

Das, S.R., Jose, J., Samal, M.R., Zhang, S., Panwar, N., 2021. Testing the star formation scaling relations in the clumps of the North American and Pelican nebulae cloud complex. MNRAS 500 (3), 3123–3141. http://dx.doi.org/10.1093/mnras/staa3222, arXiv:2010.07333.

Deharveng, L., Zavagno, A., Caplan, J., 2005. Triggered massive-star formation on the borders of Galactic H II regions. I. A search for "collect and collapse" candidates.

A&A 433 (2), 565–577. http://dx.doi.org/10.1051/0004-6361:20041946, arXiv:astro-ph/0412602.

Elmegreen, B., 2011. Triggered star formation. EAS Publ. Ser. 51, 45–58. http://dx.doi.org/10.1051/eas/1151004.

Enokiya, R., Fukui, Y., 2021. Evidence for the Major Cloud-Cloud Collisions in the Central Molecular Zone. In: Tsuboi, M., Oka, T. (Eds.), New Horizons in Galactic Center Astronomy and beyond. In: Astronomical Society of the Pacific Conference Series, vol. 528, p. 279.

Espaillat, C., Muzerolle, J., Najita, J., Andrews, S., Zhu, Z., Calvet, N., Kraus, S., Hashimoto, J., Kraus, A., D'Alessio, P., 2014. An Observational Perspective of Transitional Disks. In: Beuther, H., Klessen, R.S., Dullemond, C.P., Henning, T. (Eds.), Protostars and Planets VI. pp. 497–520. http://dx.doi.org/10.2458/azu_uapress_9780816531240-ch022, arXiv:1402.7103.

Fang, M., Hillenbrand, L.A., Kim, J.S., Findeisen, K., Herczeg, G.J., Carpenter, J.M., Rebull, L.M., Wang, H., 2020. The first extensive spectroscopic study of Young stars in the north america and pelican nebulae. Astrophys. J. 904 (2), 146. http://dx.doi.org/10.3847/1538-4357/abba84.

Froebrich, D., Scholz, A., Campbell-White, J., Crumpton, J., D'Arcy, E., Makin, S.V., Zegmott, T., Billington, S.J., Hibbert, R., Newport, R.J., Fisher, C.R., 2018. Variability in IC5070: Two Young Stars with Deep Recurring Eclipses. Res. Notes Am. Astronom. Society 2 (2), 61. http://dx.doi.org/10.3847/2515-5172/aacd48, arXiv:1806.06608.

Fujita, S., Tsutsumi, D., Ohama, A., Habe, A., Sakre, N., Okawa, K., Kohno, M., Hattori, Y., Nishimura, A., Torii, K., Sano, H., Tachihara, K., Kimura, K., Ogawa, H., Fukui, Y., 2021. High-mass star formation in Orion possibly triggered by cloud-cloud collision. III. NGC 2068 and NGC 2071. PASJ 73, S273–S284. http://dx.doi.org/10.1093/pasj/psaa005, arXiv:1706.05664.

Fukui, Y., Torii, K., Hattori, Y., Nishimura, A., Ohama, A., Shimajiri, Y., Shima, K., Habe, A., Sano, H., Kohno, M., Yamamoto, H., Tachihara, K., Onishi, T., 2018. A New Look at the Molecular Gas in M42 and M43: Possible Evidence for Cloud-Cloud Collision that Triggered Formation of the OB Stars in the Orion Nebula Cluster. ApJ 859 (2), 166. http://dx.doi.org/10.3847/1538-4357/aac217, arXiv:1701.04669.

Ge, Q.A., Li, J.J., Hao, C.J., Lin, Z.H., Hou, L.G., Liu, D.J., Li, Y.J., Bian, S.B., 2024. Evolution of the local spiral structure revealed by OB-type stars in gaia DR3. Astron. J. 168 (1), 25. http://dx.doi.org/10.3847/1538-3881/ad5201.

Guarcello, M.G., Micela, G., Peres, G., Prisinzano, L., Sciortino, S., 2010. Chronology of star formation and disk evolution in the Eagle Nebula. A&A 521, A61. http://dx.doi.org/10.1051/0004-6361/201014351, arXiv:1008.0422.

Herbert, C., Froebrich, D., Vanaverbeke, S., Scholz, A., Eislöffel, J., Urtly, T., Walton, I.L., Wiersema, K., Quinn, N.J., Piehler, G., Aimar, M.M., García, R.C., Vanmunster, T., Alfaro, F.C.S., de la Cuesta, F.G., Licchelli, D., Perez, A.E., Fernández Mañanes, E., Graciá Ribes, J., González, J.L.S., Futcher, S.R.L., Nelson, T., Dvorak, J., Moździerski, D., Kotysz, K., Mikołajczyk, P., Fleming, M., Phillips, M., Vale, T., Dubois, F., Eggenstein, H.-B., Heald, M.A., Lewin, P., OKeeffe, D., Popowicz, A., Bernacki, K., Malcher, A., Lasota, S., Fiolka, J., Dustor, A., Percy, S.C., Devine, P., Patel, A.L., Dickers, M.D., Dover, L., Grozdanova, I.I., Urquhart, J.S., Lynch, C.J.R., 2024. A survey for variable young stars with small telescopes - IX. Evolution of spot properties on YSOs in IC 5070. MNRAS 529 (4), 4856–4878. http://dx.doi.org/10.1093/mnras/stae812, arXiv:2403.10595.

Herczeg, G.J., Hillenbrand, L.A., 2014. An Optical Spectroscopic Study of T Tauri Stars. I. Photospheric Properties. ApJ 786 (2), 97. http://dx.doi.org/10.1088/0004-637X/786/2/97, arXiv:1403.1675.

Hillenbrand, L.A., Kiker, T.J., Gee, M., Lester, O., Braunfeld, N.L., Rebull, L.M., Kuhn, M.A., 2022. A Zwicky Transient Facility Look at Optical Variability of Young Stellar Objects in the North America and Pelican Nebulae Complex. AJ 163 (6), 263. http://dx.doi.org/10.3847/1538-3881/ac62d8, arXiv:2203.09633.

Indebetouw, R., Robitaille, T.P., Whitney, B.A., Churchwell, E., Babler, B., Meade, M., Watson, C., Wolfire, M., 2007. Embedded Star Formation in the Eagle Nebula with Spitzer GLIMPSE. ApJ 666 (1), 321–338. http://dx.doi.org/10.1086/520316, arXiv:0707.1895.

Jose, J., 2014. Stellar Content and Star Formation in Young Clusters Influenced by Massive Stars. In: Massive Young Star Clusters Near and Far: From the Milky Way To Reionization. pp. 47–50.

Jose, J., Pandey, A.K., Ogura, K., Bhatt, B.C., 2012. Young stellar population in the star-forming complex Sh2-252. In: Astronomical Society of India Conference Series. In: Astronomical Society of India Conference Series, vol. 4, p. 91.

Kahle, K.A., Wyrowski, F., König, C., Christensen, I.B., Tiwari, M., Menten, K.M., 2024. The effects of stellar feedback on molecular clumps in the Lagoon Nebula (M8). A&A 687, A162. http://dx.doi.org/10.1051/0004-6361/202349009, arXiv:2404.07920.

Kerton, C.R., Arvidsson, K., Knee, L.B.G., Brunt, C., 2008. Sequential and spontaneous star formation around the mid-infrared halo H ii region KR 140. Mon. Not. R. Astron. Soc. (ISSN: 0035-8711) 385 (2), 995–1002. http://dx.doi.org/10.1111/j.1365-2966.2008.12895.x.

Kinoshita, S.W., Nakamura, F., Wu, B., 2021. Star formation triggered by shocks. Astrophys. J. 921 (2), 150. http://dx.doi.org/10.3847/1538-4357/ac1d4b.

Koenig, X.P., Leisawitz, D.T., Benford, D.J., Rebull, L.M., Padgett, D.L., Assef, R.J., 2012. Wide-field Infrared Survey Explorer Observations of the Evolution of Massive Star-forming Regions. ApJ 744 (2), 130. http://dx.doi.org/10.1088/0004-637X/744/2/130.

Kounkel, M., Stassun, K.G., Covey, K., Hartmann, L., 2022. A gravitational and dynamical framework of star formation: the Orion nebula. MNRAS 517 (1), 161–174. http://dx.doi.org/10.1093/mnras/stac2695, arXiv:2111.01159.

Kuhn, M.A., Hillenbrand, L.A., Carpenter, J.M., Avelar Menendez, A.R., 2020. The formation of a stellar association in the NGC 7000/IC 5070 complex: Results from kinematic analysis of stars and gas. Astrophys. J. 899 (2), 128. http://dx.doi.org/10.3847/1538-4357/aba19a.

Lada, C.J., Forbrich, J., Lombardi, M., Alves, J.F., 2012. Star Formation Rates in Molecular Clouds and the Nature of the Extragalactic Scaling Relations. ApJ 745 (2), 190. http://dx.doi.org/10.1088/0004-637X/745/2/190, arXiv:1112.4466.

Li, S., Frank, A., Blackman, E.G., 2014. Triggered star formation and its consequences. Mon. Not. R. Astron. Soc. (ISSN: 0035-8711) 444 (3), 2884–2892. http://dx.doi.org/10.1093/mnras/stu1571.

Lozinskaya, T.A., Larkina, V.V., Putilina, E.V., 1983. A New Search for Ring Nebulae around Of-Stars - SHARPLESS22. Sov. Astron. Lett. 9, 344.

MacFarlane, B., Stamatellos, D., Johnstone, D., Herczeg, G., Baek, G., Chen, H.-R.V., Kang, S.-J., Lee, J.-E., 2019. Observational signatures of outbursting protostars - I: From hydrodynamic simulations to observations. MNRAS 487 (4), 5106–5117. http://dx.doi.org/10.1093/mnras/stz1512, arXiv:1906.01960.

Maíz Apellániz, J., Sota, A., Arias, J.I., Barbá, R.H., Walborn, N.R., Simón-Díaz, S., Negueruela, I., Marco, A., Leão, J.R.S., Herrero, A., Gamen, R.C., Alfaro, E.J., 2016. The Galactic O-Star Spectroscopic Survey (GOSSS). III. 142 Additional O-type Systems.. ApJS 224 (1), 4. http://dx.doi.org/10.3847/0067-0049/224/1/4, arXiv:1602.01336.

Megier, A., Strobel, A., Galazutdinov, G.A., Krełowski, J., 2009. The interstellar Ca II distance scale. A&A 507 (2), 833–840. http://dx.doi.org/10.1051/0004-6361/20079144.

Morris, M., Serabyn, E., 1996. The Galactic Center Environment. ARAA 34, 645–702. http://dx.doi.org/10.1146/annurev.astro.34.1.645.

NASA, JPL-Caltech, E., Hurt, R., 2013. Id: eso1339e. artist's impression of the milky way (updated - annotated).

Nishimura, A., Costes, J., Inaba, T., Tachihara, K., Hattori, Y., Kohno, M., Ohama, A., Torii, K., Sano, H., Yamamoto, H., Hasegawa, Y., Kimura, K., Ogawa, H., Fukui, Y., 2017. A new view of the giant molecular cloud M16 (Eagle Nebula) in 12CO J=1-0 and 2-1 transitions with NANTEN2. http://dx.doi.org/10.48550/arXiv.1706.06002, ArXiv E-Prints. arXiv:1706.06002.

Ohama, A., Tsutsumi, D., Sano, H., Torii, K., Nishimura, A., Shima, K., Asao Hiroaki Yamamoto, H., Tachihara, K., Hasagawa, Y., Kimura, K., Ogawa, H., Fukui, Y., 2017. High-mass star formation in Orion triggered by cloud-cloud collision II, Two merging molecular clouds in NGC2024. http://dx.doi.org/10.48550/arXiv.1706.05652, ArXiv E-Prints. arXiv:1706.05652.

Panwar, N., Jose, J., Rishi, C., 2023. Survey of hα emission-line stars in the star-forming region ic 5070. J. Astrophys. Astron. 44 (1), 42. http://dx.doi.org/10.1007/s12036-023-09935-x, arXiv:2302.06115.

Preibisch, T., Zinnecker, H., 2007. Sequentially triggered star formation in OB associations. In: Elmegreen, B.G., Palous, J. (Eds.), Triggered Star Formation in a Turbulent ISM. 237, pp. 270–277. http://dx.doi.org/10.1017/S1743921307001597, arXiv:astro-ph/0610826.

Reid, M.J., Menten, K.M., Zheng, X.W., Brunthaler, A., Moscadelli, L., Xu, Y., Zhang, B., Sato, M., Honma, M., Hirota, T., Hachisuka, K., Choi, Y.K., Moellenbrock, G.A., Bartkiewicz, A., 2009. Trigonometric Parallaxes of Massive Star-Forming Regions. VI. Galactic Structure, Fundamental Parameters, and Noncircular Motions. ApJ 700 (1), 137–148. http://dx.doi.org/10.1088/0004-637X/700/1/137, arXiv:0902.3913.

Roman-Duval, J., Heyer, M., Brunt, C.M., Clark, P., Klessen, R., Shetty, R., 2016. Distribution and Mass of Diffuse and Dense CO Gas in the Milky Way. ApJ 818 (2), 144. http://dx.doi.org/10.3847/0004-637X/818/2/144, arXiv:1601.00937.

Russeil, D., 2003. Star-forming complexes and the spiral structure of our Galaxy. A&A 397, 133–146. http://dx.doi.org/10.1051/0004-6361:20021504.

Sabatini, G., Bovino, S., Giannetti, A., Grassi, T., Brand, J., Schisano, E., Wyrowski, F., Leurini, S., Menten, K.M., 2021. Establishing the evolutionary timescales of the massive star formation process through chemistry. A&A 652, A71. http://dx.doi.org/10.1051/0004-6361/202140469, arXiv:2106.00692.

Schmeja, S., Klessen, R.S., Froebrich, D., 2005. Number ratios of young stellar objects in embedded clusters. A&A 437 (3), 911–918. http://dx.doi.org/10.1051/0004-6361:20041898, arXiv:astro-ph/0503611.

Sharpless, S., 1959. A Catalogue of H II Regions.. ApJS 4, 257. http://dx.doi.org/10.1086/190049.

Sicilia-Aguilar, A., Kahar, R.S., Pelayo-Baldárrago, M.E., Roccatagliata, V., Froebrich, D., Galindo-Guil, F.J., Campbell-White, J., Kim, J.S., Mendigutía, I., Schlueter, L., Teixeira, P.S., Matsumura, S., Fang, M., Scholz, A., Ábrahám, P., Frasca, A., Garufi, A., Herbert, C., Kóspál, Á., Manara, C.F., 2024. North-PHASE: studying periodicity, hot spots, accretion stability, and early evolution in young stars in the Northern hemisphere. MNRAS 532 (2), 2108–2132. http://dx.doi.org/10.1093/mnras/stae1588, arXiv:2406.16702.

Snider, K.D., Hester, J.J., Desch, S.J., Healy, K.R., Bally, J., 2009. Spitzer Observations of The H II Region NGC 2467: An Analysis of Triggered Star Formation. ApJ 700 (1), 506–522. http://dx.doi.org/10.1088/0004-637X/700/1/506, arXiv:0711.1515.

Sugitani, K., Tamura, M., Nakajima, Y., Nagashima, C., Nagayama, T., Nakaya, H., Pickles, A.J., Nagata, T., Sato, S., Fukuda, N., Ogura, K., 2002. Near-Infrared Study of M16: Star Formation in the Elephant Trunks. ApJL 565 (1), L25–L28. http://dx.doi.org/10.1086/339196.

Tiwari, M., Menten, K.M., Wyrowski, F., Giannetti, A., Lee, M.Y., Kim, W.J., Pérez-Beaupuits, J.P., 2020. Cause and effects of the massive star formation in Messier 8 East. A&A 644, A25. http://dx.doi.org/10.1051/0004-6361/202038886, arXiv:2010.09365.

Tothill, N.F.H., Gagné, M., Stecklum, B., Kenworthy, M.A., 2008. The Lagoon Nebula and its Vicinity. In: Reipurth, B. (Ed.), Handbook of Star Forming Regions, Volume II. vol. 5, p. 533. http://dx.doi.org/10.48550/arXiv.0809.3380.

Urquhart, J.S., Figura, C.C., Moore, T.J.T., Hoare, M.G., Lumsden, S.L., Mottram, J.C., Thompson, M.A., Oudmaijer, R.D., 2014. The RMS survey: galactic distribution of massive star formation. MNRAS 437 (2), 1791–1807. http://dx.doi.org/10.1093/mnras/stt2006, arXiv:1310.4758.

van der Marel, N., 2023. Transition disks: the observational revolution from SEDs to imaging. Eur. Phys. J. Plus 138 (3), 225. http://dx.doi.org/10.1140/epjp/s13360-022-03628-0, arXiv:2210.05539.

Vázquez-Semadeni, E., Zamora-Avilés, M., Galván-Madrid, R., Forbrich, J., 2018. Molecular cloud evolution - VI. Measuring cloud ages. MNRAS 479 (3), 3254–3263. http://dx.doi.org/10.1093/mnras/sty1586, arXiv:1805.07221.

Walker, D.L., Longmore, S.N., Bally, J., Ginsburg, A., Kruijssen, J.M.D., Zhang, Q., Henshaw, J.D., Lu, X., Alves, J., Barnes, A.T., Battersby, C., Beuther, H., Contreras, Y.A., Gómez, L., Ho, L.C., Jackson, J.M., Kauffmann, J., Mills, E.A.C., Pillai, T., 2021. Star formation in 'the Brick': ALMA reveals an active protocluster in the Galactic centre cloud G0.253+0.016. MNRAS 503 (1), 77–95. http://dx.doi.org/10.1093/mnras/stab415, arXiv:2102.03560.

Wright, N.J., 2020. OB Associations and their origins. New Astronomy Reviews 90, 101549. http://dx.doi.org/10.1016/j.newar.2020.101549, arXiv:2011.09483.

Wright, E.L., Eisenhardt, P.R.M., Mainzer, A.K., Ressler, M.E., Cutri, R.M., Jarrett, T., Kirkpatrick, J.D., Padgett, D., McMillan, R.S., Skrutskie, M., Stanford, S.A., Cohen, M., Walker, R.G., Mather, J.C., Leisawitz, D., Gautier, III, T.N., McLean, I., Benford, D., Lonsdale, C.J., Blain, A., Mendez, B., Irace, W.R., Duval, V., Liu, F., Royer, D., Heinrichsen, I., Howard, J., Shannon, M., Kendall, M., Walsh, A.L., Larsen, M., Cardon, J.G., Schick, S., Schwalm, M., Abid, M., Fabinsky, B., Naes, L., Tsai, C.-W., 2010. The Wide-field Infrared Survey Explorer (WISE): Mission Description and Initial On-orbit Performance. AJ 140 (6), 1868–1881. http://dx.doi.org/10.1088/0004-6256/140/6/1868, arXiv:1008.0031.

Xu, J.-L., Zavagno, A., Yu, N., Liu, X.-L., Xu, Y., Yuan, J., Zhang, C.-P., Zhang, S.-J., Zhang, G.-Y., Ning, C.-C., Ju, B.-G., 2019. The effects of ionization feedback on star formation: a case study of the M 16 H II region. A&A 627, A27. http://dx.doi.org/10.1051/0004-6361/201935024, arXiv:1905.08030.

Yadav, R.K., Pandey, A., Sharma, S., Jose, J., Ogura, K., Kobayashi, N., Samal, M., Eswaraiah, C., Chandola, H., 2015. Deep optical survey of the stellar content of Sh2-311 region. New Astron. (ISSN: 1384-1076) 34, 27–40. http://dx.doi.org/10.1016/j.newast.2014.05.004.

Yadav, R.K., Pandey, A.K., Sharma, S., Ojha, D.K., Samal, M.R., Mallick, K.K., Jose, J., Ogura, K., Richichi, A., Irawati, P., Kobayashi, N., Eswaraiah, C., 2016. A multiwavelength investigation of the H II region S311: young stellar population and star formation. MNRAS 461 (3), 2502–2518. http://dx.doi.org/10.1093/mnras/stw1356.

Yamada, R.I., Enokiya, R., Sano, H., Fujita, S., Kohno, M., Tsutsumi, D., Nishimura, A., Tachihara, K., Fukui, Y., 2021. A kinematic analysis of the CO clouds toward a reflection nebula NGC 2023 observed using the Nobeyama 45 m telescope: Further evidence for a cloud-cloud collision in the Orion region. PASJ 73 (4), 880–893. http://dx.doi.org/10.1093/pasj/psab050, arXiv:2006.03426.